\documentclass[draftcls,onecolumn,12pt]{IEEEtran}
\usepackage{bm}
\usepackage{epsfig}
\usepackage{stfloats}
\usepackage{graphicx}
\usepackage{subfigure}

\usepackage{epstopdf}
\usepackage{cite}
\usepackage{epsfig,graphics,subfigure}
\usepackage[cmex10]{amsmath}

\usepackage{array}
\usepackage{tabulary}
\usepackage{booktabs}

\begin{document}

\bibliographystyle{IEEEtran}

\title{Secrecy Outage Performance of Multi-antenna Wiretap Channels With Diversity Combinings Over Correlated Rayleigh Fading Channels}
\author{\IEEEauthorblockN{Jiangbo Si, {\emph{Member, IEEE}}, Zan Li, {\emph{Senior Member, IEEE}}, \\
Julian Cheng, {\emph{Senior Member, IEEE}}, and Caijun Zhong, {\emph{Senior Member, IEEE}}}
\thanks{}
\thanks{Jiangbo Si and Zan Li are with the Integrated Service Networks Lab
of Xidian University, Xi'an, China (e-mail: jbsi@xidian.edu.cn).}
\thanks{Julian Cheng is with the School of Engineering, The University
of British Columbia, Kelowna, BC V1V 1V7, Canada (e-mail: julian.cheng@ubc.ca).}
\thanks{Caijun Zhong is with Information and Communication Engineering, Zhejiang
University, Hangzhou 310027, China (e-mail: caijunzhong@zju.edu.cn).}
}
 \maketitle
\begin{abstract}
This paper presents a detailed secrecy outage performance analysis of correlated multi-antenna wiretap channels with three popular diversity combining schemes, namely maximal ratio combining, selection combining, and equal gain combining at the legitimate receiver. For single input multiple output wiretap channels, both exact and asymptotic expressions are derived for the secrecy outage probability (SOP) of these systems. The findings suggest that, compared with the scenario where all the channels are independent, correlation of the main channels alone increases the SOP by a factor of $1/\det(\boldsymbol{U})$, where $\det(\boldsymbol{U})$ is the determinant of correlation matrix $\boldsymbol{U}$. In contrast, in the high signal-to-noise ratio regime at the legitimate receiver, correlation between the main channels and the eavesdropper channels has a positive effect on SOP. Moreover, for multiple input and multiple output wiretap channels, the SOP of two transmit antenna selection (TAS) schemes are analyzed for the considered correlated channel model. The SOP relationship of the both TAS schemes is quantified by the secrecy array gain. Finally, numerical simulations are conducted to corroborate the analytical results.

\end{abstract}
\IEEEpeerreviewmaketitle
\section{Introduction}
With the rapid penetration of wireless services into various aspects of the society, security has become an increasing critical concern. Responding to this concern, physical layer security (PLS) technique has emerged as a promising solution to provide secrecy communication, and has gained tremendous research attention in recent years. Since the pioneer work of Bloch \emph{et al}. \cite{Bloch}, who demonstrated that positive secrecy rate can be achieved regardless of the channel quality of the legitimate link, different approaches have been proposed in the literature to further enhance the secrecy performance of PLS. Among which, multiple antenna technology can provide the diversity gain and is of particular interest.

A plethora of works have investigated the impact of multiple antennas on the secrecy performance. For instance, the secrecy outage probability (SOP) of single-input multiple-output (SIMO) and multiple-input single-output (MISO) wiretap channels was studied in \cite{HLei} and \cite{XZhou}, respectively. Later on, the secrecy performance of multiple-input multiple-output (MIMO) wiretap channels was investigated in \cite{FSAlQahtani3},\cite{YHuang}. In these prior works, it is commonly assumed that the channels are independent, which is a reasonable assumption when these multiple antennas are spaced sufficiently apart. In practice, due to space constraint, these multiple antennas can be correlated. In addition, the main channel and the eavesdropper channel can also be correlated when the eavesdropper is close to the legitimate receiver.


Therefore, understanding the impact of correlation on the secrecy performance of multi-antenna wiretap channels is of both theoretical and practical interests. By considering correlation at the multi-antenna legitimate receiver only, the effect of antenna correlation on the secrecy performance was investigated in \cite{JHu}, while the case with correlation at the multi-antenna eavesdropper only was examined in \cite{VUPrabhu}. In \cite{MZl}, the impact of equal antenna correlation at both the multi-antenna legitimate receiver and eavesdropper on SOP was studied with maximum ratio combining (MRC). Later on, the extension to the arbitrary antenna correlation case was considered in \cite{NSFerdinand} and \cite{KPPeppas}. Moreover, with antenna correlation at both the legitimate receiver and the eavesdropper, the SOP of transmit antenna selection (TAS) scheme was studied in \cite{NYang}.

In addition to the above works, which only focused on the antenna correlation, several works also investigated the secrecy performance when the main channel and eavesdropper channel are correlated. The secrecy capacity bounds for the correlated main and the eavesdropper channels were derived in \cite{HJeon}. The authors in \cite{XSun} studied the the secrecy performance over correlated fading channels in the presence of a single-antenna eavesdropper. The secrecy performance over correlated Nakagami-fading was recently presented in \cite{GCAlexandropoulos}. In \cite{NSFerdinand1}, it was shown that the correlation beteween the legitimate receiver and eavesdropper channels can benefit secrecy performance in the high signal-to-noise ratio (SNR) regime. Similarly, in the cooperative relaying systems, it was shown that correlation between the relay-destination link and the relay-eavesdropper link is beneficial for the SOP \cite{LFan}. Later on, the impacts of correlation on the SOP for different relay selection schemes were investigated in \cite{LFan1}, where the authors showed that correlation can improve the SOP for full relay selection and degrade the SOP for the partial relay selection. For cognitive networks, the impact of correlation between the main channels and the eavesdropper channels on SOP was investigated in \cite{JZhang}. Also, secrecy key generation was proposed to take advantage of correltion between the main and eavesdropper channels \cite{BHe1}.

However, one common limitation of the forementioned works is that they only considered the scenario with correlation between single antenna receiver and single antenna eavesdropper channels. In addition, the exact SOP gains due to correlation were not characterized. In modern communication system, communication terminals are likely equipped with multiple antennas. Therefore it is necessary to investigate the impact of correlation on SOP over multi-antenna wiretap channels. To the best of the authors' knowledge, the secrecy performance of general multi-antenna wiretap channels with arbitrary correlation between the main and eavesdropper's channels has not been well understood. Moreover, it is well known that the performance of diversity reception over independent channels is better than that over correlated channels \cite{MSlim}. However whether the secrecy performance over correlated channels is superior or inferor to that over independent channels is still an open question. Motivated by these, in this paper, we consider a more comprehensive system model with generalized correlaton model. We systematically study the effect of correlation on the SOP for the SIMO wiretap channels, and the relationship of SOP over independent channeles and correlated channels is quantified by simple and meaningful expressions. Furthermore, for a multi-anenna eavesdropper, we investigate the SOP of TAS schemes for the MIMO wiretap channels. Specifically, the main contributions of the paper are summarized as follows:

\begin{itemize}
\item[1.] A joint  probability density function (PDF) for a sum of correlated variables pertaining to our system model is first derived. Since the traditional correlation matrix eigenvalue decomposition technique \cite{AAlmradi} can not be used for SOP analysis in our scenario, we resort to the correlation channel model presented in \cite{Chen}-\cite{JSchlenker} to establish the correlation between the main channels and the evaesdropper channels. All correlated channels are modeled as a set of conditional independent channel gains. For this, the SOP can be derived by the method of moment generation function and conditional probability density function.

\item[2.] For three popular diversity combining schemes: MRC, selection combining (SC), and equal gain combining (EGC), simple asymptotic expressions are presented for the SOP of the considered system. The result indicates that the correlation does not change the secrecy diversity order. Moreover, the analytical findings suggest that, compared with the ideal case where all the channels are independent, antenna correlation of the main channels (CMC) increases the SOP by a factor of $1/\det(\boldsymbol{U})$ in the high SNR regime, where $\det(\boldsymbol{U})$ is the determinant of the main channels correlation matrix $\boldsymbol{U}$. In contrast, the correlation between the main and eavesdropper channels (CMEC) is beneficial for SOP. When the average SNR at the legitimate receiver is larger than that at the eavesdropper, the larger the correlation coefficient is, the more SOP gains can be obtained. Moreover, depending on the correlation coefficients, the combined effects due to correlation can either degrade or improve the SOP performance.

\item[3.] For correlated MIMO wiretap channels, two different types of TAS schemes are studied, depending on the availability of eavesdropper's channel state information (CSI). For both cases, $N_E$ antennas are deployed at the eavesdropper, and the asymptotic SOP expressions of all three different diversity combining schemes are presented. The findings suggest that the considered TAS schemes achieve the same secrecy diversity order of $MN_t$, where $N_t$ is the number of transmit antennas and $M$ is the number of the receiver antennas. According to the secrecy array gain, the SOP gap of two TAS schemes is quantified by a closed-form expression and increases with $N_t$. Especially, when $N_E=1$, compared with case of TAS with eavesdropper's CSI, the SOP of the TAS without eavesdropper's CSI deteriorates by a factor of $\frac{{\left( {MN_t } \right)!}}{{\left( {M!} \right)^{N_t } }}$.
\end{itemize}

The remainder of this paper is organized as follows.
Section II describes the system model. In Section III, assuming a single antenna at the transmitter, the asymptotic expressions of SOP are derived. In Section IV, when multiple antennas are deployed at the transmitter and the eavesdropper, the asymptotic SOP expressions of TAS are derived with or without eavesdropper's CSI.
Simulation results are presented in Section V. We conclude the paper in Section VI.

\emph{Notations}- $\Pr \left(  \cdot  \right)$ denotes the probability of
an event; $f_\beta  \left(  \cdot  \right)$ is the PDF of  random variable (RV) $\beta$; $F_\beta  \left(  \cdot  \right)$ is the cumulative distribution
function (CDF) of RV $\beta$; $I_0(\cdot)$ is the modified Bessel function of the first kind of
order zero; ${\phi _\beta }\left( s \right)$ is the moment generation function (MGF) of $\beta$; $E{(\cdot)}$ is expected value of a RV; ${\mathcal{L}^{-1}}{(\cdot)}$ denotes the inverse Laplace transform; $\det{(\cdot)}$ is the determinant of a matrix; $\Gamma{(\cdot)}$ is the Gamma function; An
$m \times m$ identity matrix is denoted by ${\bf{I}}_{m}$.  $ _1F_1(\alpha;\gamma;z)$ denotes the confluent hypergeometric function. $ {}_2F_1(\alpha,\beta;\gamma;z)$ is the hypergeometric function. $E_n(\cdot)$ is the exponential-integral function. For a vector $\boldsymbol{x}$, ${\boldsymbol{x}}^T$ denotes the transpose, and ${\boldsymbol{x}}^H$ denotes the conjugate transpose.
\section{System Model}
\begin{figure}[ht]
\centering
\includegraphics[width=2.4in]{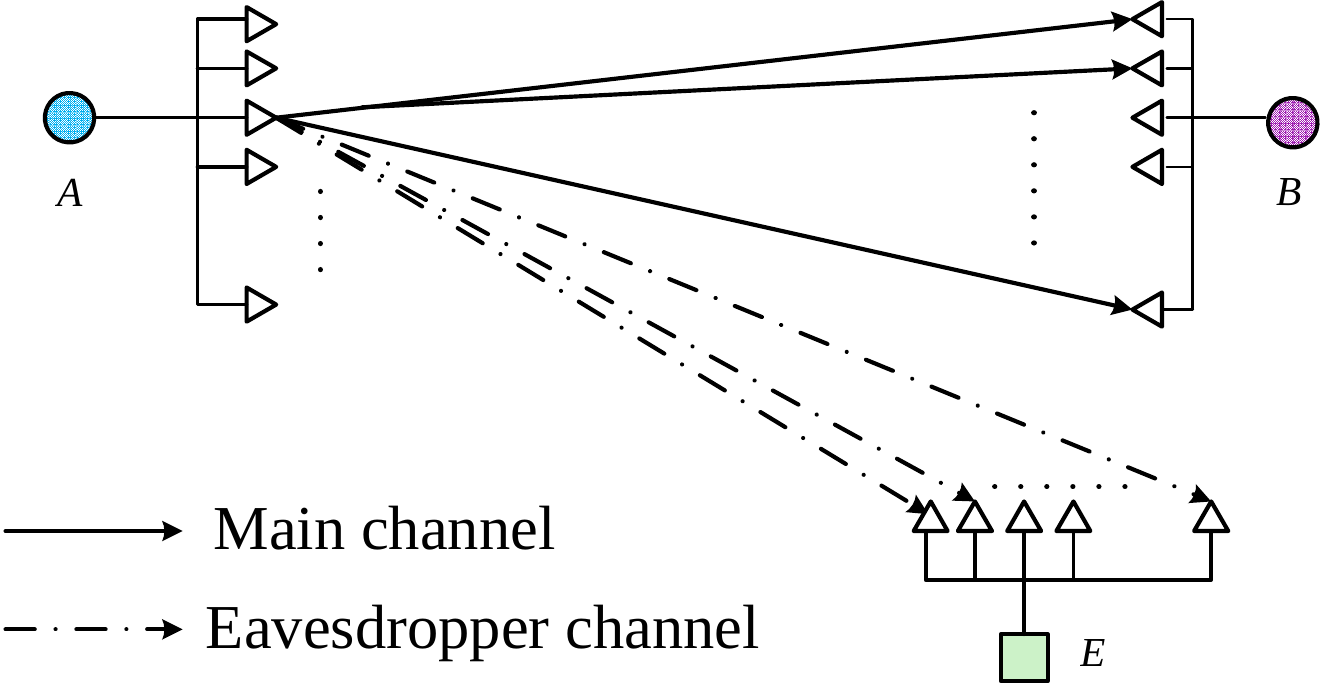}
\caption{System model of a multi-antenna secure communication systems} \label{fig.1}
\end{figure}
Consider a MIMO wiretap channel illustrated in Fig. 1, where the Alice or source node $S$ equipped with $N_t$ antennas aims to establish a secure communication link with the legitimate receiver $B$ equipped with $M$ antennas in the presence of eavesdropper $E$ equipped with $N_E$ antennas. All the channels are assumed to experience Rayleigh fading. When the signal is transmitted by the $n$-th antenna, the received signal at the legitimate receiver can be expressed as
\begin{align}\label{27}\
{\bf{y}}_{s_n, B}  = \sqrt{P_0} {\bf{h}}_{s_n,B} x + {\bf{n}}_B ,  {n = 1, \cdots, {N_t}}
\end{align}
where $x$ is the information symbol with unit power, $P_0$ is the source transmit power, and ${\bf{h}}_{s_n,B}  = \left\{ {{{h}}_{s_n,b_1 } ,{{h}}_{s_n,b_2 } , \cdots, {{h}}_{s_n,b_M } } \right\}^T$ is an $M\times1$ channel vector between the $n$-th transmitter antenna and receiver. Also, ${{h}}_{s_n,b_m }$ $(1\leq m \leq M)$ is a complex Gaussian RV with mean zero and variance $\frac{{\sigma ^2 _{s,b } }}{2}$. Finally, the $M\times 1$ vector ${\bf{n}}_B$ denotes the additive white Gaussian noise (AWGN) at legitimate receiver satisfying $E\left( {{\bf{n}}_B {\bf{n}}^H _B } \right) = {\bf{I}}_M N_0$, where $N_0$ is the noise variance. Since the legitimate receiver has multiple antennas, diversity combining techniques can be employed to enhance the secrecy performance. Depending on the specific diversity combining schemes, the SNR at legitimate receiver, $\gamma_{s_n,B}$, takes different forms, which will be specified in the ensuing section. Similarly, the received signal at the eavesdroppers can be expressed as
\begin{align}\label{27}\
{\bf{y}}_{s_n,E}  = \sqrt{P_0} {\bf{h}}_{s_n,E} x + {\bf{n}}_E ,  {n = 1, \cdots, {N_t}}
\end{align}
where ${\bf{h}}_{s_n,E}  = \left\{ {{{h}}_{s_n,e_1 } ,{{h}}_{s_n,e_2 } , \cdots, {{h}}_{s_n,e_{N_E} } } \right\}^T$ is the $N_E\times1$ channel vector between the $n$-th transmitter antenna and the eavesdropper;  ${{h}}_{s_n,e_k }$ $(1\leq k \leq N_E)$ is an AWGN RV with mean zero and variance $\frac{{\sigma ^2 _{s,e} }}{2}$; ${\bf{n}}_E$ is an AWGN RV vector at the eavesdropper satisfying $E\left( {{\bf{n}}_B {\bf{n}}^H _B } \right) = {\bf{I}}_{N_E} N_0$.

When multiple antennas are deployed at the legitimate receiver, due to the space constraint, the channels at multiple antennas are assumed to be correlated. In \cite{Chen} and \cite{XZhang}, the authors noted that a set of related complex Gaussian RVs can be obtained by linearly combining a set of independent Gaussian RVs.  Then the correlated channel between the $n$-th antenna at the transmitter and the $m$-th antenna at the legitimate receiver, $h_{s_n,b_m}$, can be expressed as \cite{JSchlenker}
\begin{align}\label{correlaionmodelb}\
 h_{s_n,b_m}  &= {\sigma _{s_n,b_m}} \left[ {\left( {\sqrt {1 - \eta^2_{m} } X_{s_n,b_m}  +  {\eta_{m} } X_{n,0} } \right)} \right. \nonumber \\
  & {\left. { + i\left( {\sqrt {1 - \eta^2_{m} } Y_{s_n, b_m}  +  {\eta_{m} } Y_{n,0} } \right)} \right]}, {}{} m = 1, \cdots, M
\end{align}
where $i=\sqrt{-1}$, $-1\leq\eta_{m}\leq1$, and $X_{s_n,b_m}$, $Y_{s_n,b_m}$, $X_{n,0}$, $Y_{n,0}$ are independent Gaussian RVs with distribution ${\cal{N}}(0,1/2)$. For any ${m},{{m'}}\in\{1,...M \}$, $
{E}\left( {X_{s_n,b_m} X_{s_n,b_{m'}} } \right) = {E}\left( {Y_{s_n,b_m} Y_{s_n,b_{m'}} } \right) = \frac{1}{2}\delta _{m,{m'}}
$, where $\delta_{u,v}$ is the Kronecker delta function, i.e., $\delta _{m,{m'}}  =1$ for ${m = {m'}}$ and $\delta _{m,{m'}}  =1$ for ${m \neq {m'}}$. Since the distance between the transmit antenna and the receive antenna is far greater than that between any two transimt antennas in practice, we assume all tranmit antennas at the legitimate receiver have the same correlation matrix,  which is defined as
\begin{equation}\boldsymbol{U}=\left(\begin{array}{cccc}
1 & {{\rho^{b} _{1,2}}}  & \cdots &  {{\rho^{b} _{1,M}}} \\
 {{\rho^{b} _{1,2}}}  & 1 & \cdots &  {{\rho^{b} _{2,M}}} \\
\vdots & \vdots & \ddots & \vdots \\
 {{\rho^{b} _{1,M}}} &  {{\rho^{b} _{2,M}}} & \cdots & 1
 \end{array}\right)\end{equation}
where the element $\rho^{b} _{m,m'}$ $(1\leq m \leq M, 1\leq m' \leq M , m \neq m' )$ denotes the correlation coefficient between the $m$-th and $m'$-th receive antenna at the legitimate receiver, and it is calculated as
\begin{align}\label{correlationcoefficient}\
\rho^{b} _{m,{m'}}  = \frac{{E\left( {h_{s_n,b_m} ,h^*_{s_n,b_{m'}} } \right) - E\left( {h_{s_n,b_m} } \right)E\left( {h^*_{s_n,b_{m'}} } \right)}}{{\sqrt {{{E}}\left( {|h_{s_n ,b_m}|^2 } \right){{E}}\left( |{h_{s_n,b_{m'}} }|^2 \right)} }} = {\eta _m \eta_{m'} }
\end{align}

When the eavesdropper is close to the legitimate receiver, the signal received at the legitimate receiver can be correlated with signal at the eavesdropper. Also, when the eavesdropper lies in the radio wave path of legitimated signal and the eavesdropper is less than 70-80 wavelengths from the legitimate receiver (see \cite{HJeon} and reference therein), the received signal at the legitimate receiver and the eavesdropper can be correlated. Thus, we can model the eavesdropper's channel as
\begin{align}\label{correlaionmodele}\
 h_{s_n,e_k} &={\sigma _{s,e}} \left[ {\left( {\sqrt {1 - \lambda^2_{k} } X_{s_n,e_k}  +  {\lambda_{k} } X_{n,0} } \right)} \right. \nonumber  \\
         & {\left. { + i\left( {\sqrt {1 - \lambda^2_{k} } Y_{s_n,e_k}  + {\lambda_{k} } Y_{n, 0} } \right)} \right]}, {}{} k = 1, \cdots, {N_E}
\end{align}
where $-1\leq\lambda_{k}\leq1$; $X_{s_n,e_k}$ and $Y_{s_n, e_k}$ are independent Gaussian RVs with distribution ${\cal{N}}(0,1/2)$. In this paper, for analytical convenience, we assume the eavesdropper channels are equally correlated, $\lambda_{k} =\lambda_e$. Thus, similar to \eqref{correlationcoefficient}, the correlation coefficient between ${{{h_{s_n, e_{\vartheta}}}} }$ and ${ {{h_{ s_n, e_{{\vartheta}'}}}}}$ $(1\leq \vartheta \leq N_E)$ is
\begin{align}\label{27}\
\rho^{e}_{\vartheta,{{\vartheta}'}} = {\lambda^2 _e}.
\end{align}

As an important contribution of this work, we establish correlaion between the main channels and the eavesdropper channels via the common variables $X_{n,0}$ and $Y_{n,0}$ in both \eqref{correlaionmodelb} and \eqref{correlaionmodele}.  As a result, the correlation coefficient between ${{{h_{s_n, b_m}}} }$ and ${ {{h_{ s_n, e_\vartheta}}}}$ can be concisely expressed as
\begin{align}\label{27}\
\rho^{b,e}_{m,\vartheta} = {\eta_m \lambda_e }.
\end{align}

 To evaluate the secrecy performance for more general scenarios, classical
information-theoretic methods are often deployed \cite{HLei}-\cite{JZhang}. For the information-theoretic secrecy evaluation, the SOP, non-zero capacity rate, and average secrecy capacity are three popular and common metrics \footnote{It is worth noting that several new secrecy performance metrics have recently been proposed by taking into account of the eavesdropper's decodability of confidential messages \cite{BHe}.}\cite{Bloch}. Since we consider the case that the eavesdropper's CSI is unavailable, and the
probability of non-zero secrecy rate can be easily obtained by the special case of SOP, we will employ the SOP to evaluate the secrecy performance. When the $n$-th transmit antenna is selected, the secrecy capacity is expressed as
\begin{align}\label{27}\
C_{s_n}  = \left\{ \begin{array}{l}
 {\begin{array}{*{20}c}
 \log _2 \left( {1 + \gamma _{s_n,B} } \right) - \log _2 \left( {1 + \gamma _{s_n,E} } \right), &{\gamma _{s_n,B}  \ge \gamma _{s_n,E}} \\
 \end{array}} \\
   {\begin{array}{*{20}c}
   0, \quad\quad \quad \quad \quad \quad \quad \quad \quad \quad \quad \quad \quad \quad \quad  &{\gamma _{s_n,B} < \gamma _{s_n,E}} \\
\end{array}} \\
 \end{array} \right.
\end{align}
where $\gamma _{s_n,B}$ and $\gamma _{s_n,E}$ denote the SNR at the legitimate receiver and eavesdropper, respectively. The SOP is defined as the probability that $C_{s_n}$ is less than a predetermined secrecy rate $R_s$. It can be written as \cite{LFan2}, \cite{FSAlQahtani}
\begin{align}\label{sopdefinition}\
 P_{out} \left( {R_s } \right) &= \Pr \left( {C_{s_n}  < R_s } \right) \nonumber \\
  &= \Pr \left( {\gamma _{s_n, B}  < 2^{R_s } \left( {1 + \gamma _{s_n, E} } \right)-1} \right).
\end{align}
In the remainder of the paper, the definition in \eqref{sopdefinition} is used to evaluate the secrecy performance.
\section{SOP With single antenna at the transmitter and the eavesdropper}
In this section, we analyze the SOP of a SIMO system, which may correspond to an uplink of a cellular network, where the base station is equipped with multiple antennas, and the terminal is often equipped with single antenna due to size and cost constraint. It is valuable to evaluate the SOP performance of the SIMO system with a single-antenna eavesdropper, per the discussions in \cite{XSun}-\cite{LFan1}. Moreover, the case of multi-antenna eavesdopper will be considered in Section IV. This section also forms the basis of analysis for TAS with or without eavesdropper's CSI in Section IV.

In particular, we consider three different diversity combining schemes adopted at the legitimate receiver, namely, MRC, SC and EGC. Each combining scheme has its advantage and disadvantage. MRC requires accurate instantaneous CSI and can achieve the best performance. SC compares all branches instantaneous CSI and has a simple structure. However, the performance of SC is much lower than that of MRC. To compromise the system performance and complexity, EGC is a suitable choice. For the correlated SIMO wiretap channels, we consider these three popular diversity combining schemes at the legitimate receiver. Next, the impacts of correlation on SOP for three combining schemes are analytically quantified in detail.
\subsection{MRC at the Legitimate Receiver}
When MRC is adopted at the legitimate receiver, the SNR is given by
 \begin{align}\label{snrMRC}\
\gamma^{MRC}_{s_n,B}  = \gamma _{s_n,b_1 }  + \gamma _{s_n,b_2 }  +  \cdots  + \gamma _{s_n,b_M }
\end{align}
where ${\gamma _{s_n,{b_m}}} = \frac{{{P_0}{{\left| {{h_{s_n,{b_m}}}} \right|}^2}}}{{{N_0}}}$ ($1 \le m \le M$) is exponentially distributed with mean $\bar\gamma_{s_n,b_m}
=\frac{{P_0 \sigma ^2 _{s_n,b_m} }}{{N_0 }}$. Since the antennas are co-located, it is reasonable to assume that the main channels at the legitimate receiver have the same average SNR $\bar\gamma_{s_n,b_m}=\bar\gamma_{B}$.
\subsubsection{Exact SOP}
When the main channels and the eavesdropper channel are correlated. The SNR at the destination has been presented in (\ref{snrMRC}). Since $\gamma^{MRC}_{s_n,B}$  and $\gamma_{s_n, E}$ are correlated, and all the variables in (\ref{snrMRC}) and $\gamma_{s_n, E}$ are correlated, it is challenging to derive the exact SOP for arbitrarily correlated fading channels. We only discuss the exact SOP for equally correlated case. We can resort to the conditional CDF to derive the exact SOP. We assume ${T=X^2_{n,0}+Y^2_{n, 0}}$, $X_{n,0}=x_0$, and $Y_{n,0}=y_0$. If the main channels are equally correlated, i. e., $\rho^{b}_{m,{m'}}=\rho$, conditioned on ${{x}}^{{2}} _0  + {{y}}^{{2}} _0  = t$, $\gamma^{MRC}_{s_n,B}$ has a noncentral chi-square distribution  $\chi _{{2M}} \left( {\sqrt {M\bar\gamma_{B} \rho^2 t}, \frac{{\bar\gamma _{B} \left( {1 - \rho^2 } \right)}}{2}} \right)$, and the conditional CDF of $\gamma^{MRC}_{s_n,B}$ is given by
\begin{align}\label{eqconditianl}\
F_{\gamma^{MRC}_{s_n,B}|T} \left( {x\left| t \right.} \right) = 1 - Q_M \left( {\sqrt {\frac{{2M\rho^2 t}}{{1 - \rho^2}}} ,\sqrt {\frac{{2x}}{{\bar\gamma _B \left( {1 - \rho^2} \right)}}} } \right)
\end{align}
where $Q_M \left( {\cdot, \cdot} \right)$ denotes the $M$-th order Marcum
$Q$-function, and it is defined as \cite{Chen}
\begin{align}\label{Qfunction}\
Q_M \left( {a,b} \right) = \int_b^\infty  {x\left( {\frac{x}{a}} \right)} ^{M - 1} \exp \left( { - \left( {\frac{{x^2  + a^2 }}{2}} \right)} \right)I_{M - 1} \left( {ax} \right)dx.
\end{align}
Similarly, conditioned on ${{x}}^{{2}} _0  + {{y}}^{{2}} _0  = t$, $\gamma _{s_n,E}$ has a noncentral chi-square distribution \\$\chi _{{2}} \left( {\sqrt {\bar\gamma_{E} \lambda ^2_{{e}}t} ,\frac{{\bar\gamma _{E} \left( {1 - \lambda^2_e } \right)}}{2}} \right)$, and the conditional CDF of $\gamma_{s_n, E}$ is given by
\begin{align}\label{condtionalCDFre}\
F_{{\gamma _{s_n, E} }|T} \left( {y\left| t \right.} \right) = 1 - Q_1\left( {\sqrt {\frac{{2\lambda^2_e t}}{{1 - \lambda^2_e }}} ,\sqrt {\frac{{2y}}{{\bar\gamma _E \left( {1 - \lambda^2_e } \right)}}} } \right)
\end{align}
where $Q_1(\cdot,\cdot)$ denotes the first-order Marcum $Q$-function. Differentiating (\ref{condtionalCDFre}) with respect to $y$, the conditional PDF of $\gamma_{s_n, E}$ is derived as
\begin{align}\label{conpdfre}\
{f_{{\gamma _{s_n, E}}|T}}\left( {y\left| t \right.} \right) = \frac{{\exp \left( { - \frac{{{\lambda^2_e}t}}{{1 - {\lambda^2}}}} \right)}}{{{\bar\gamma _E}\left( {1 - {\lambda^2_e}} \right)}}\exp \left( { - \frac{y}{{{\bar\gamma _E}\left( {1 - {\lambda^2_e}} \right)}}} \right){I_0}\left( {2\sqrt {\frac{{{\lambda^2_e}t y}}{\bar\gamma_E({1 - {\lambda^2_e}})^2}}} \right).
\end{align}
According to (\ref{sopdefinition}), (\ref{eqconditianl}), and (\ref{conpdfre}), the SOP of MRC is given by
\begin{align}\label{eqMRC}\
P_{out\_MRC}\left( {{R_s}} \right) &= \int_0^\infty  {\exp \left( { - t} \right)} \Pr \left( {{\gamma^{MRC}_{s_n,B}} \le {2^{{R_s}}}\left( {{\gamma _{s_n, E}} + 1} \right) - 1\left| t \right.} \right)dt \nonumber \\
 &= \int_0^\infty  {{e^{ - t}}dt} \int_0^\infty  {\left( {1 - {Q_M}\left( {\sqrt {\frac{{2M{\rho^2}t}}{{1 - {\rho^2}}}} ,\sqrt {\frac{{2\left( {{2^{{R_s}}}\left( {y + 1} \right) - 1} \right)}}{{{\bar\gamma _B}\left( {1 - {\rho^2}} \right)}}} } \right)} \right)} \nonumber \\
 &\times \frac{{\exp \left( { - \frac{{{\lambda^2_e}t}}{{1 - {\lambda^2}}}} \right)}}{{{\bar\gamma _E}\left( {1 - {\lambda^2_e}} \right)}}\exp \left( { - \frac{y}{{{\bar\gamma _E}\left( {1 - {\lambda^2_e}} \right)}}} \right){I_0}\left( {2\sqrt {\frac{{{\lambda^2_e}ty}}{{{\bar\gamma _E}{{\left( {1 - {\lambda^2_e}} \right)}^2}}}} } \right)dy.
\end{align}
By using the function MARCUMQ provided in Matlab, the exact SOP of MRC can be obtained.
\subsubsection{Asymptotic SOP}  If the correlation coefficients are arbitrary, it is challenging to obtain the conditional CDF or PDF of $\gamma^{MRC}_{s_n,B}$ and the direct method to derive the SOP is intractable. We resort to the conditional MGF to obtain the asymptotic SOP with arbitrary correlation at the legitimate receiver.  As $\bar\gamma _B \rightarrow \infty$, it corresponds to the scenario where the main
channels have good quality while the eavesdropper's channel is severely blocked due to severe shadowing.  The asymptotic conditional PDF of $\gamma^{MRC}_{s_n,B}$ is derived in Appendix A. According to (\ref{conpdfre}) and (\ref{asyconpdf}), as $\bar\gamma _B \rightarrow \infty$, we can obtain the asymptotic SOP of MRC as
\begin{align}\label{asyallcMRC}\
&\mathop {P_{out\_MRC}}\limits_{{{\bar \gamma }_B} \to \infty } \left( {{R_s}} \right) = \int_0^\infty  {\exp \left( { - t} \right)} dt\int_0^\infty  {\left( {{\gamma^{MRC}_{s_n,B}} \le {2^{{R_s}}}\left( {1 + y} \right) - 1} \right){f_{{\gamma _{s_n,E}}}}\left( {y\left| t \right.} \right)} dy  \nonumber \\
&\quad= \frac{{{{\bar \gamma }_E}^M}}{{\det \left(\boldsymbol{U} \right){{\bar \gamma }_B}^M}}\sum\limits_{k = 0}^M {\frac{1}{{\left( {M - k} \right)!}}{\left( {\frac{{1 + \sum\limits_{m = 1}^M {\frac{{{\eta^2_m}\left( {1 - {\lambda^2_e}} \right)}}{{1 - {\eta^2_m}}}} }}{{1 + \sum\limits_{m = 1}^M {\frac{{{\eta^2_m}}}{{1 - {\eta^2_m}}}} }}} \right)^k}{{\left( {\frac{{{2^{{R_s}}} - 1}}{{{{\bar \gamma }_E}}}} \right)}^{M - k}}} {\left( {{2^{{R_s}}}} \right)^k}+ o\left( {\frac{1}{{\bar \gamma _B ^{M } }}} \right)
\end{align}
where in obtaining the last equality the following integral identity
\begin{align}\label{formula}\
\int_0^\infty  {\exp \left( { - \mu x} \right)} I_0 \left( {2\sqrt {\upsilon x} } \right)dx = \frac{1}{\mu }\exp \left( {\frac{\upsilon }{\mu }} \right)
\end{align}
is used \cite{IS}. In (\ref{asyallcMRC}), $o(\cdot) $ denotes higher order
terms; $\det{\boldsymbol{(U)}}$ is the determinant of matrix $\boldsymbol{U}$, and it is given by \cite{CDMeyer}
\begin{align}\label{deltc}
\det \left( \boldsymbol {U} \right) = \prod\limits_{m = 1}^M {\left( {1 - {\eta^2_m}} \right)\left( {1 + \sum\limits_{m = 1}^M {\frac{{{\eta^2_m}}}{{1 - {\eta^2_m}}}} } \right).}
\end{align}
It can be seen from (\ref{asyallcMRC}) that the diversity order is $M$. In addition, if $\bar\gamma_{E}\gg 1$, eq. (\ref{asyallcMRC}) can be further simplified as
\begin{align}\label{formula}\
P_{out\_MRC} \left( {R_s } \right) \simeq \frac{1}{{\det \left( \boldsymbol{U} \right)}}\left( {\frac{{2^{R_s } \bar \gamma _E }}{{\bar \gamma _B }}} \right)^M \left( {\frac{{1 + \sum\limits_{m = 1}^M {\frac{{\eta ^2 _m \left( {1 - \lambda^2_e } \right)}}{{1 - \eta ^2 _m }}} }}{{1 + \sum\limits_{m = 1}^M {\frac{{\eta ^2 _m }}{{1 - \eta ^2 _m }}} }}} \right)^M.
\end{align}
Since the asymptotic SOP with $\bar\gamma_{E}\gg 1$ can be easily obtained by the aysmptotic SOP, in the remainder of the paper, unless otherwise stated, we will not derive the asymptotic SOPs with $\bar\gamma_{E}\gg 1$ .
\subsection{SC at the Legitimate Receiver}
When the SC is applied, the SNR at the multi-antennas legitimate receiver is given by
\begin{align}\label{SNRsc}\
{\gamma^{SC} _{s_n,B}} = \max \left( {{\gamma _{s_n,{b_1}}},{\gamma _{s_n,{b_2}}}, \cdots {\gamma _{s_n,{b_M}}}} \right).
\end{align}
\subsubsection{Exact SOP} When SC is applied at the legitimate receiver, the SNR at the legitimate receiver has been given in (\ref{SNRsc}). Since the main and eavesdropper's channels are correlated, similar to the case of MRC, the conditional CDF approach is used. For fixed $X_{n,0}=x_0$, $Y_{n,0}=y_0$, the conditional ${{{h_{ s_n, b_m}}} }$ is complex Gaussian distributed with mean ${ \sqrt {\bar\gamma _{B}\eta^2_m } \left( {x_0  + iy_0 } \right)}$ and variance ${\frac{{\bar\gamma _{B}\left( {1 - \eta^2_m } \right)}}{2}}$. ${{|{h_{s_n, b_m}}|^2} }$ obeys a noncentral chi-square distribution $\chi _{{2}} \left( {\sqrt {\bar\gamma_{B} \eta^2 _{{m}} \left( {{{x}}^{{2}} _0  + {{y}}^{{2}} _0 } \right)} ,\frac{{\bar\gamma _{B} \left( {1 - \eta^2_m } \right)}}{2}} \right)$. Assuming ${T=X^2_{n,0}+Y^2_{n, 0}}$ with ${{x}}^{{2}} _0  + {{y}}^{{2}} _0  = t$, the conditional CDF of ${{{\gamma_{s_n, b_m}}} }$ can be obtained as
\begin{align}\label{CDFsbk}\
{F_{\left. {{\gamma _{s_n,{b_m}}}} \right|T}}\left( {\left. x \right|t} \right) &= \Pr \left( {{{\gamma_{s_n, b_m}}}   \le x\left| t \right.} \right) \nonumber \\
 &= 1 - Q_1\left( {\sqrt {\frac{{2\eta^2_{{m}} t}}{{{1 - \eta^2_m } }}} ,\sqrt {\frac{{2x}}{{\bar\gamma _{B} \left( {1 - \eta^2_m } \right)}}} } \right).
\end{align}
According to (\ref{SNRsc}) and (\ref{PDFsbk}), the conditional CDF of ${\gamma^{SC} _{s_n,B}}$ is expressed as
\begin{align}\label{fsc}\
F_{\gamma^{SC} _{s_n,B}|T } \left( x|t \right) =  {\prod\limits_{m = 1}^M {\left( {1 - Q_1\left( {\sqrt {\frac{{2\eta^2_m t}}{{1 - \eta^2_m }}} ,\sqrt {\frac{{2x}}{{\bar \gamma _B \left( {1 - \eta^2_m } \right)}}} } \right)} \right)} }.
\end{align}
The SOP of SC is given by
\begin{align}\label{exactsop}
&P_{out\_SC}\left( {{R_s}} \right)= \int_0^\infty  {{e^{ - t}}\Pr \left( {{\gamma^{SC} _{s_n,B}} \le \left( {{2^{{R_s}}}\left( {{\gamma _{s_n,E}} + 1} \right) - 1} \right)\left| t \right.} \right)dt}\nonumber \\
 &= \int_0^\infty  {{e^{ - t}}dt} \int_0^\infty  {\prod\limits_{m = 1}^M {Q_1\left( {\sqrt {\frac{{2{\eta^2_m}t}}{{1 - {\eta^2_m}}}} ,\sqrt {\frac{{2\left( {{2^{{R_s}}}\left( {y + 1} \right) - 1} \right)}}{{{\bar\gamma _B}\left( {1 - {\eta^2_m}} \right)}}} } \right)} {f_{{\gamma _{s_n,E}}}}\left( {y\left| t \right.} \right)} dy \nonumber\\
 &= \int_0^\infty  {\prod\limits_{m = 1}^M {Q_1\left( {\sqrt {\frac{{2{\eta^2_m}t}}{{1 - {\eta^2_m}}}} ,\sqrt {\frac{{2\left( {{2^{{R_s}}}\left( {y + 1} \right) - 1} \right)}}{{{\bar\gamma _B}\left( {1 - {\eta^2_m}} \right)}}} } \right)} \exp \left( { - \frac{y}{{{\bar\gamma _E}\left( {1 - {\lambda^2_e}} \right)}}} \right){I_0}\left( {\sqrt {\frac{{2{\lambda^2_e}t}}{{1 - {\lambda^2_e}}}} y} \right)} dy \nonumber \\
 &\quad \times \int_0^\infty  {\frac{{\exp \left( { - \frac{t}{{1 - {\lambda^2_e}}}} \right)}}{{{\bar\gamma _B}\left( {1 - {\lambda^2_e}} \right)}}} dt.
\end{align}
\subsubsection{Asymptotic SOP}  As $\beta\rightarrow 0^{+}$, $Q_M(\alpha ,\beta)$  can be approximated as \cite{ChengJun2Mar}
 \begin{align}\label{asyQfunction}
 {Q_M}\left( {\alpha ,\beta } \right) \approx 1 - \frac{{{\beta ^{2M}}}}{{{2^M}M!}}\exp \left( { - \frac{{{\alpha ^2}}}{2}} \right).
\end{align}
As $\bar\gamma_{B}\rightarrow\infty$, substituting (\ref{asyQfunction}) into (\ref{exactsop}), the asymptotic SOP is derived as
\begin{align}\label{asyallcsc}\
\mathop {P_{out\_SC}}\limits_{\bar \gamma _B  \to \infty} \left( {{R_s}} \right) = \frac{{M!{{\bar \gamma }_E}^M}}{{\det \left( \boldsymbol{U} \right){{\bar \gamma }_B}^M}}\sum\limits_{k = 0}^M {\frac{{\left( {{2^{{R_s}}}} \right)^k}}{{\left( {M - k} \right)!}}} {\left( {\frac{{1 + \sum\limits_{k = 1}^M {\frac{{{\eta^2_m}\left( {1 - {\lambda^2_e}} \right)}}{{1 - {\eta^2_m}}}} }}{{1 + \sum\limits_{m = 1}^M {\frac{{{\eta^2_m}}}{{1 - {\eta^2_m}}}} }}} \right)^k}{{\left( {\frac{{{2^{{R_s}}} - 1}}{{{{\bar \gamma }_E}}}} \right)}^{M - k}}+ o\left( {\frac{1}{{\bar \gamma _B ^{M } }}} \right).
\end{align}
It can be seen from (\ref{asyallcsc}) that the diversity order is still $M$.
\subsection{EGC at the Legitimate Receiver}
When the EGC is applied, the received signals are first co-phased and then summed to form the resultant output. The SNR at the multi-antenna legitimate receiver is given by
\begin{align}\label{SNREGC}
\gamma^{EGC} _{s_n, B}  = \frac{{P_0 \left( {\sum\limits_{m = 1}^M {\left| {h_{s_n,b_m } } \right|} } \right)^2 }}{{MN_0 }}
 = \frac{{\left( {\sum\limits_{m = 1}^M { \sqrt{\gamma_{s_n,b_m } } } } \right)^2 }}{M}.
\end{align}
Since $\gamma^{EGC} _{s_n, B}$ is correlated with $\gamma _{E}$, it is challenging to derive $\gamma^{EGC} _{s_n, B}$ directly. Different from the case of MRC, we derive the conditional PDF of $\gamma^{EGC} _{s_n, B}$ in Appendix B. Then according to (\ref{sopdefinition}), (\ref{conpdfre}), and (\ref{PDFZ2}), the asymptotic SOP of EGC is derived as
\begin{align}\label{asyegc}\
 &\mathop {P _{out\_EGC} \left( {R_s } \right)}\limits_{\bar \gamma _B  \to \infty }= \Pr \left( {\gamma^{EGC} _{s_n, B}  \le \left( {2^{R_s } \left( {\gamma _{s_n,E}  + 1} \right) - 1} \right)} \right)  \nonumber  \\
  &= \int_0^\infty  {e^{ - t} dt\int_0^\infty  {f_{\left. {\gamma _{s_n,E} } \right|T} \left( {\left. y \right|t} \right)} \int_0^{2^{R_s } \left( {y + 1} \right) - 1} {f_{\left. {\gamma^{EGC} _{s_n, B} } \right|T} } \left( {\left. x \right|t} \right)dx} dy \nonumber  \\
  &=\frac{{M!\left( {2M} \right)^M \bar \gamma _E ^M }}{{\left( {2M} \right)!\det \left( \boldsymbol{U} \right)\bar \gamma _B ^M }}\sum\limits_{k = 0}^M {\frac{1}{{\left( {M - k} \right)!}}\left( {\frac{{1 + \sum\limits_{m = 1}^M {\frac{{\eta^2_m \left( {1 - \lambda^2_e } \right)}}{{1 - \eta^2_m }}} }}{{1 + \sum\limits_{m = 1}^M {\frac{{\eta^2_m }}{{1 - \eta^2_m }}} }}} \right)^k \left( {\frac{{2^{R_s }  - 1}}{{\bar \gamma _E }}} \right)^{M - k} \left( {2^{R_s } } \right)^k }+ o\left( {\frac{1}{{\bar \gamma _B ^{M } }}} \right).
\end{align}
It can be observed from (\ref{asyegc}) that the diversity order is still $M$.
\subsection{Special Case of $\lambda_e=0$}
As $\lambda_e=0$, the main channels and the eavesdropper channel are independent. Eqs. (\ref{asyallcMRC}), (\ref{asyallcsc}), and (\ref{asyegc}) are, respectively, simplified to
\begin{align}\label{sopMRCasy}
\mathop {P^{C}_{out\_MRC}}\limits_{{{\bar \gamma }_B} \to \infty } \left( {{R_s}} \right)& = \frac{{{{ {{{\bar \gamma }_E}}}^M}}}{{{{ {{{\bar \gamma }_B}} }^M}\det \left( \boldsymbol{U} \right)}}\sum\limits_{k = 0}^M {\frac{1}{{\left( {M - k} \right)!}}{{\left( {\frac{{{2^{{R_s}}} - 1}}{{{{\bar \gamma }_E}}}} \right)}^{M - k}}} {\left( {{2^{{R_s}}}} \right)^k}+ o\left( {\frac{1}{{\bar \gamma _B ^{M } }}} \right)
\end{align}
\begin{align}\label{sopscasy}
\mathop {{P^{C}_{out\_SC}}}\limits_{{{\bar \gamma }_B} \to \infty } \left( {{R_s}} \right) &=\frac{{M!{{{{{\bar \gamma }_E}}}^M}}}{{{{ {{{\bar \gamma }_B}}}^M}\det \left(\boldsymbol{U} \right)}}\sum\limits_{k = 0}^M {\frac{1}{{\left( {M - k} \right)!}}{{\left( {\frac{{{2^{{R_s}}} - 1}}{{{{\bar \gamma }_E}}}} \right)}^{M - k}}} {\left( {{2^{{R_s}}}} \right)^k}+ o\left( {\frac{1}{{\bar \gamma _B ^{M } }}} \right)
\end{align}
\begin{align}\label{sopegcasy}\
\mathop {P^{C} _{out\_EGC} \left( {R_s } \right)}\limits_{\bar \gamma _B  \to \infty } =  \frac{{M!\left({2M} \right)^M\bar \gamma _E ^M }}{{\left( {2M} \right)!\det \left( \boldsymbol{U} \right)\bar \gamma _B ^M }}\sum\limits_{k = 0}^M {\frac{1}{{\left( {M - k} \right)!}}\left( {\frac{{2^{R_s }  - 1}}{{\bar \gamma _E }}} \right)^{M - k} \left( {2^{R_s } } \right)^k }+ o\left( {\frac{1}{{\bar \gamma _B ^{M } }}} \right).
\end{align}
\paragraph{$\eta^2_m=1$} When $\eta^2_{m}=1$ $(1\leq m \leq M)$ or $\det{(\boldsymbol{U})}=0$, the main channels are completely correlated. In this case,  the SOP of different combining schemes is given by
\begin{align}\label{completecorrelatedasy}
\mathop {P^{C} _{out} }\limits_{\scriptstyle \eta^2_m  = 1 \hfill \atop
  \scriptstyle\bar \gamma _B  \to \infty  \hfill} (R_s) =\frac{{\bar \gamma _E }}{{\varepsilon\bar \gamma _B }}\left( {2^{R_s }  + \frac{{2^{R_s }  - 1}}{{\bar \gamma _E  }}} \right).
\end{align}
In \eqref{completecorrelatedasy}, as MRC and EGC are deployed at the legitimate receiver, $\varepsilon = M$. As SC is deployed, $\varepsilon = 1$. It is obvious from (\ref{completecorrelatedasy}), the secrecy diversity order decreases to unity when the main channels are completely correlated.
\paragraph{$\eta_m=0$} When $\lambda_e=0$, and $\eta_m=0$, namely $\det({\boldsymbol{U}})=1$, all the channels are independent. Eqs. (\ref{asyallcMRC}), (\ref{asyallcsc}), and (\ref{asyegc}) are, respectively, simplified as
\begin{align}\label{asysopMRCindependent}
\mathop {P^I _{out\_MRC} }\limits_{\bar \gamma _B  \to \infty } \left( {R_s } \right)= \frac{{{{ {{{\bar \gamma }_E}} }^M}}}{{{{ {{{\bar \gamma }_B}} }^M}}}\sum\limits_{k = 0}^M {\frac{1}{{\left( {M - k} \right)!}}{{\left( {\frac{{{2^{{R_s}}} - 1}}{{{{\bar \gamma }_E}}}} \right)}^{M - m}}} {\left( {{2^{{R_s}}}} \right)^k}+ o\left( {\frac{1}{{\bar \gamma _B ^{M } }}} \right)
\end{align}
 \begin{align}\label{independtsopscasy}
\mathop {{P^{I}_{out\_{SC}}}}\limits_{{{\bar \gamma }_B} \to \infty } \left( {{R_s}} \right) =  \frac{{M!{{{{{\bar \gamma }_E}}}^M}}}{{{{ {{{\bar \gamma }_B}}}^M}}}\sum\limits_{k = 0}^M {\frac{1}{{\left( {M - k} \right)!}}{{\left( {\frac{{{2^{{R_s}}} - 1}}{{{{\bar \gamma }_E}}}} \right)}^{M - k}}} {\left( {{2^{{R_s}}}} \right)^k}+ o\left( {\frac{1}{{\bar \gamma _B ^{M } }}} \right)
\end{align}
\begin{align}\label{independtsopegcasy}\
\mathop {P^{I} _{out\_EGC} \left( {R_s } \right)}\limits_{\bar \gamma _B  \to \infty } &= \frac{{M!\left( {2M} \right)^M \bar \gamma _E ^M }}{{\left( {2M} \right)!\bar \gamma _B ^M }}\sum\limits_{k = 0}^M {\frac{1}{{\left( {M - k} \right)!}}\left( {\frac{{2^{R_s }  - 1}}{{\bar \gamma _E }}} \right)^{M - k} \left( {2^{R_s } } \right)^k }+ o\left( {\frac{1}{{\bar \gamma _B ^{M } }}} \right).
\end{align}

\emph{Remark 1}: Comparing the case of having only antenna correlation at the legitimate receiver with the case where all the channels are independent, we reveal that antenna correlation at the legitimate receiver increases the SOP by a factor of $1/\det(\boldsymbol{U})$. Since ${\det \left( \boldsymbol{U}  \right)}\leq 1$ \cite{SLiu}, \cite{XSong}, antenna correlation at the legitimate receiver deteriorates the SOP. The higher antenna correlation becomes, the worse SOP is obtained. Moreover, it can be seen that regardless which diversity reception is deployed at the legitimate receiver, CMC does not change the secrecy diversity order $M$. In addition, it can be seen that as $\bar \gamma _B  \to \infty$, no matter whether the main channels are correlated or not, the SOP of SC is $M!$ times that of MRC, and the SOP of EGC is  $\frac{M!{\left( {2M} \right)^M }}{{\left( {2M} \right)!}}$ times that of MRC.
\subsection{Comparison of CMEC With CMC } Comparing (\ref{asyallcMRC}), (\ref{asyallcsc}), (\ref{asyegc}) with(\ref{sopMRCasy}),(\ref{sopscasy}), (\ref{sopegcasy}),  as $\bar\gamma_{E}\gg 1$, for the three combining schemes we can find
\begin{align}\label{ratioMRC}\
\mathop {\lim }\limits_{{{\bar \gamma }_B} \to \infty } \frac{{P_{out}\left( {{R_s}} \right)}}{P^{C}_{out}\left( {{R_s}} \right)} = {\left( {\frac{{1 + \sum\limits_{k = 1}^M {\frac{{{\eta^2_m}\left( {1 - {\lambda^2_e}} \right)}}{{1 - {\eta^2_m}}}} }}{{1 + \sum\limits_{k = 1}^M {\frac{{{\eta^2_m}}}{{1 - {\eta^2_m}}}} }}} \right)^M}.
\end{align}
Because $ {\frac{{1 + \sum\limits_{m= 1}^M {\frac{{{\eta^2_m}\left( {1 - {\lambda^2_e}} \right)}}{{1 - {\eta^2_m}}}} }}{{1 + \sum\limits_{m = 1}^M {\frac{{{\eta^2_m}}}{{1 - {\eta^2_m}}}} }}} \leq 1$, $P_{out\_MRC}\left( {{R_s}} \right) \le P^{C}_{out\_MRC}\left( {{R_s}} \right)$, $P_{out\_SC}\left( {{R_s}} \right) \le P^{C}_{out\_SC}\left( {{R_s}} \right)$, and $P_{out\_EGC}\left( {{R_s}} \right) \le P^{C}_{out\_EGC}\left( {{R_s}} \right)$ .

\emph{Remark 2:}  we can conclude that in the high SNR regime at the legitimate receiver, the correlation between the main channels and the eavesdropper channel improves the SOP performance. While similar conclusion was made for the single-antenna main channel and the single-antenna eavesdropper channel \cite{NSFerdinand1}. We emphasize that the same conclusion can be made for multi-anntenna main channels here, as well as for multi-antenna evesdropper channels in Section IV. This phenomenon can be intuitively understood by analyzing the SOP under the extreme case that the main and eavesdropper channels are fully correlated, where the instantaneous SNR at the legitimate receiver is always larger than that at the eavesdropper, and the best SOP performance can be obtained. Note that \eqref{ratioMRC} is obtained when $\bar\gamma_B$ approaches infinity. Hence, our statement holds under the condition that the average SNR $\bar\gamma_B$ at the legitimate receiver is much larger than the average SNR at the eavesdropper $\bar\gamma_E$.
\subsection{Comparison of CMEC with Independent of All The Channels} Comparing (\ref{asyallcMRC}) with  (\ref{asysopMRCindependent}), (\ref{asyallcsc}) with (\ref{independtsopscasy}), and (\ref{asyegc}) with (\ref{independtsopegcasy}), as $\bar\gamma_{E}\gg 1$, for the three combining schemes, the relationship between the SOP under the case when all the channels are correlated and the SOP under the case when all the channels are independent is given by
\begin{align}\label{ratioMRC1}\
\mathop {\lim }\limits_{{{\bar \gamma }_B} \to \infty } \frac{{P_{out}\left( {{R_s}} \right)}}{{P^I_{out}\left( {{R_s}} \right)}} =  \frac{1}{{\det \left( \boldsymbol{U}  \right)}}{\left( {\frac{{1 + \sum\limits_{m = 1}^M {\frac{{{\eta^2_m}\left( {1 - {\lambda^2_e}} \right)}}{{1 - {\eta^2_m}}}} }}{{1 + \sum\limits_{k = 1}^M {\frac{{{\eta^2_m}}}{{1 - {\eta^2_m}}}} }}} \right)^M}.
\end{align}

\emph{Remark 3}: Setting ${P^I_{out}\left( {{R_s}} \right)}$ as the reference, it can be easily concluded from \eqref{ratioMRC1} that, in the high SNR regime, the SOP decreases with $\lambda^2_e$. Moreover, note that in \eqref{ratioMRC1} the first part $\frac{1}{{\det \left( \boldsymbol{U}  \right)}}\geq 1$, and the second part ${\left( {\frac{{1 + \sum\limits_{m = 1}^M {\frac{{{\eta^2_m}\left( {1 - {\lambda^2_e}} \right)}}{{1 - {\eta^2_m}}}} }}{{1 + \sum\limits_{k = 1}^M {\frac{{{\eta^2_m}}}{{1 - {\eta^2_m}}}} }}} \right)^M}\leq 1$. This leads to $\frac{{P_{out}\left( {{R_s}} \right)}}{{P^I_{out}\left( {{R_s}} \right)}}\leq 1$ when ${\lambda^2_e}$ is large and the main channels are slightly correlated. In other words, the combined effects of the antenna correlation at the legitimate receiver ($\eta_m$) and the correlation between the main channels and eavesdropper channel ($\lambda_e$) can either degrade or improve the SOP performance.
\subsection{Lower and Upper SOP Bounds with Correlation Between The Main and Eavesdropper Channels}
In Remark 2, we have known the SOP decreases with $\lambda^2_e$, when the main and eavesdropper channels are correlated. Then with fixed antenna correlation matrix $\boldsymbol{U}$ and the fixed average channel gains $\bar\gamma_{B}$ and $\bar\gamma_{E}$, the lower bound of SOP can be obtained by letting $\lambda^2_e=1$.  For the special case of $\lambda^2_e=1$, as $\bar\gamma_{B}\rightarrow\infty$, the asymptotic SOP of MRC is recalculated as
\begin{align}\label{mrc2}\
 &P_{out\_MRC}= \Pr \left( {\gamma^{mrc} _{s_n,B}  \le 2^{R_s } \left( {\gamma _E  + 1} \right) - 1} \right)= \int_0^\infty  {e^{ - t} } \int_0^{2^{R_s } \left( {\bar \gamma _E t + 1} \right) - 1} {f_{\gamma _{B\_mrc} } \left( x \right)dx}  \nonumber \\
  &\quad=\frac{{\bar \gamma _E ^M }}{{\bar \gamma _B ^M \det \left( \boldsymbol{U} \right)}}\sum\limits_{k = 0}^M {\frac{1}{{\left( {1 + \sum\limits_{m = 1}^M {\frac{{\eta ^2 _m }}{{1 - \eta ^2 _m }}} } \right)^k }}\frac{1}{{\left( {M - k} \right)!}}\left( {\frac{{2^{R_s }  - 1}}{{\bar \gamma _E }}} \right)^{M - k} } 2^{R_s k}+ o\left( {\frac{1}{{\bar \gamma _B ^{M } }}} \right).
\end{align}
Similarly, the asymptotic SOPs of SC and EGC are rewritten as
\begin{align}\label{sc2}\
P_{out\_SC}  = \frac{{M!\bar \gamma _E ^M }}{{\bar \gamma _B ^M \det \left( \boldsymbol{U} \right)}}\sum\limits_{k = 0}^M {\frac{1}{{\left( {1 + \sum\limits_{k = 1}^M {\frac{{\eta ^2 _m }}{{1 - \eta ^2 _m }}} } \right)^k }}\frac{1}{{\left( {M - k} \right)!}}\left( {\frac{{2^{R_s }  - 1}}{{\bar \gamma _E }}} \right)^{M - k} } 2^{R_s k}+ o\left( {\frac{1}{{\bar \gamma _B ^{M } }}} \right)
\end{align}
and
\begin{align}\label{EGCc2}\
P_{out\_EGC}  =\frac{{M!\left( {2M} \right)^M \bar \gamma _E ^M }}{{\left( {2M} \right)!\bar \gamma _B ^M \det \left( \boldsymbol {U} \right)}}\sum\limits_{k = 0}^M {\frac{2^{R_s k}}{{\left( {M - k} \right)!}{\left( {1 + \sum\limits_{k = 1}^M {\frac{{\eta ^2 _m  }}{{1 - \eta ^2 _m  }}} } \right)^k }}\left( {\frac{{2^{R_s }  - 1}}{{\bar \gamma _E }}} \right)^{M - k} } + o\left( {\frac{1}{{\bar \gamma _B ^{M } }}} \right).
\end{align}
It can be observed that (\ref{mrc2}), (\ref{sc2}), and (\ref{EGCc2}) are respectively the lower SOP bounds for MRC, SC and EGC, when the main and eavesdropper channels are correlated. Moreover, these equations coincide with (\ref{asyallcMRC}), (\ref{asyallcsc}) and (\ref{asyegc}) with $\lambda^2_e=1$. Also, the upper SOP bounds correspond to the case of $\lambda^2_e=0$, and they have been obtained in \eqref{sopMRCasy}, \eqref{sopscasy}, and \eqref{sopegcasy}.
\section{SOP of TAS with the multi-antenna eavesdropper}
 Multiple antennas technique can be applied at both the base station and the terminals at the legitimate receiver and the eavesdropper. It is therefore necessary to investigate the SOP of a MIMO system. In addition, considering that the multi-antenna eavesdropper tries to overhear the legitimate information as much as possible, we only consider the case when MRC is deployed at the eavesdropper. Moverover, due to constraints on cost and complexity, a single RF chain is equipped at the transmitter. Only one transmit antenna is selected to forward the information. It is worth mentioning that the TAS technique has
been approved by IEEE 802.16 for WiMAX and
LTE-Advanced \cite{NBMehta} since it can achieve full diversity at low
cost and low implementation complexity in Rayleigh fading
channels \cite{ZChen}. Hence, under both cases of with or without eavesdropper's CSI, the asymptotic SOP expressions of TAS/MRC, TAS/SC, and TAS/EGC are derived over correlated Raleigh fading channels in the following. We comment that while the SOP for TAS with diversity receptions was studied for correlatd legitimate channels and correlated eavesdropper channels \cite{NYang}, the effects of correaltions between the multi-branch main channels and the multi-branch eavesdropper channels have not been investigated.
\subsection{Without Eavesdropper's CSI}
When the eavesdropper's CSI is unavailable to the transmitter, the transmit antenna having the maximum SNR at the legitimate receiver is selected, and it can be expressed as \cite{NYang2}
\begin{align}\label{protcol}\
n^*  = \arg \mathop {\max }\limits_{1 \le n \le N_t } {\gamma _{s_n ,B} }
\end{align}
where $\gamma _{s_n ,B}$ is determined by the specific combining scheme deployed at the legitimate receiver. When MRC, SC, and EGC are applied,  $\gamma _{s_n ,B}$ is respectively denoted by $\gamma^{MRC} _{s_n ,B}$, $\gamma^{SC} _{s_n ,B}$ and $\gamma^{EGC} _{s_n ,B}$.
\subsubsection{MRC at the Legitimate Receiver}
Note when the eavesdropper's channel is independent of the main channels, the selected antenna $n^*$ is a random antenna for eavesdropper $E$. But when the eavesdropper's channel is correlated with the main channels, the SNR at the eavesdropper $\gamma^{MRC}_{s_{n^*},E}$ is correlated with the SNR at the legitimate receiver $\gamma^{MRC}_{s_{n^*},B}$. Then conditioned on ${{x}}^{{2}} _0  + {{y}}^{{2}} _0  = t$, the SNR at the eavesdropper, $\gamma^{MRC}_{s_n,E}$, has a noncentral chi-square distribution  $\chi _{{2N_E}} \left( {\sqrt {N_E\bar\gamma_{E} \lambda^2_e t}, \frac{{\bar\gamma _{E} \left( {1 - \lambda^2_e } \right)}}{2}} \right)$, and the conditional PDF of $\gamma^{MRC}_{s_n,E}$ is given by
\begin{align}\label{pdfemrc}\
f_{\left. {\gamma^{MRC} _{s_n ,E} } \right|T} \left( {\left. y \right|t} \right) = \frac{\exp \left( { - \frac{{y + N_E\bar \gamma _E {\lambda ^2_e} t}}{{\bar \gamma _E \left( {1 - \lambda ^2_e } \right)}}} \right)}{{\bar \gamma _E \left( {1 - \lambda ^2_e } \right)}}\left( {\frac{y}{{N_E\bar \gamma _E {\lambda ^2_e} t}}} \right)^{\frac{{N_E  - 1}}{2}}I_{N_E  - 1} \left( {2\sqrt {\frac{{N_E\lambda ^2_e ty}}{{\bar \gamma _E \left( {1 - \lambda ^2_e } \right)^2 }}} } \right).
\end{align}
The conditional PDF of $\gamma^{MRC}_{s_{n^*},B}$ has been derived in \eqref{asyconpdf}. Then with the help of \eqref{pdfemrc} and \eqref{asyconpdf}, the joint PDF of $\gamma^{MRC}_{s_{n^*},B}$ and $\gamma^{MRC}_{s_{n^*},E}$ is given in the following lemma.

\emph{Lemma 1:} The joint PDF of $\gamma^{MRC}_{s_{n^*},B}$ and $\gamma^{MRC}_{s_{n^*},E}$ can be obtained as
\begin{align}\label{jointpdf}
&f_{\gamma^{MRC}_{s_{n^*,B}},\gamma^{MRC}_{s_{n^*,E}}} \left( {x,y} \right)= \left( {\frac{1}{{\Gamma \left( {M + 1} \right)\det \left( \boldsymbol{U} \right)}}} \right)^{N_t  - 1} \frac{{N_t \prod\limits_{m = 1}^M {\frac{1}{{1 - \eta ^2 _m }}\left( {\frac{1}{{1 - \lambda ^2_e }}} \right)^{N_E } } }}{{\Gamma \left( M \right)\bar \gamma _B ^{MN_t } \bar \gamma _E ^{N_E } \Gamma \left( {N_E } \right)\alpha }}x^{MN_t  - 1}  \nonumber \\
  &\quad\quad\quad\times y^{N_E  - 1} \exp \left( { - \frac{y}{{\bar \gamma _E \left( {1 - \lambda ^2_e } \right)}}} \right){}_1F_1 \left( {1;N_E ;\frac{{N_E {\lambda ^2_e} y}}{{\alpha \bar \gamma _E \left( {1 - \lambda ^2_e } \right)^2 }}} \right) + o\left( {\frac{{x^{MN_t  - 1} }}{{\bar \gamma _B ^{MN_t } }}} \right).
\end{align}
\emph{Proof:} The derivation is presented in Appendix C. According to (\ref{sopdefinition}) and (\ref{jointpdf}), the asymptotic SOP of TAS/MRC is calculated as
\begin{align}\label{multimrc}
 &P^{TAS} _{out\_MRC} \left( {R_s } \right)= \Pr \left( {\frac{{\gamma^{MRC} _{s_{n^*} ,B}  + 1}}{{\gamma^{MRC} _{s_{n^*} ,E}  + 1}} \le 2^{R_s } } \right)= \int_0^\infty  {\int_0^{2^{R_s } \left( {y + 1} \right) - 1} {f_{\gamma ^{MRC} _{s_{n^*} ,B} ,\gamma^{MRC} _{s_{n^*} ,E} } \left( {x,y} \right)} } dxdy \nonumber \\
  &= \frac{1}{{\prod\limits_{m = 1}^M {\left( {1 - \eta ^2 _m } \right)} \det \left( \boldsymbol{U} \right)^{N_t  - 1} \Gamma ^{N_t } \left( {M + 1} \right)\Gamma \left( {N_E } \right)\alpha \bar \gamma _B ^{MN_t } }}\sum\limits_{\omega  = 0}^{MN_t } {\binom{ MN_t }{\omega }} \left( {2^{R_s }  - 1} \right)^{MN_t  - \omega }  \nonumber \\
  &\times \Gamma \left( {N_E  + \omega } \right)\left( {2^{R_s } \bar \gamma _E \left( {1 - \lambda ^2_e } \right)} \right)^\omega  {}_2F_1\left( {1,N_E  + \omega ;N_E ;\frac{{N_E \lambda ^2_e }}{{\alpha \left( {1 - \lambda ^2_e } \right)}}} \right) + o\left( {\frac{1}{{\bar \gamma _B ^{MN_t } }}} \right).
\end{align}
As $\eta_m=0$, $\lambda_e=0$, $\det \left( \boldsymbol{U} \right)=1$, and all channels are independent, eq. \eqref{multimrc} is reduced to \cite[eq. (27)]{NYang2} over Rayleigh fading. This demonstrates the generality for our result. As $\bar\gamma_E\gg1$, the asymptotic SOP of TAS/MRC is rewritten as
\begin{align}\label{multimrc2}
P^{TAS} _{out\_MRC} \left( {R_s } \right) &= \frac{{\Gamma \left( {MN_t  + N_E } \right)\left( {\bar \gamma _E 2^{R_s } \left( {1 - \lambda^2_e } \right)} \right)^{MN_t } }}{{\prod\limits_{m = 1}^M {\left( {1 - \eta^2_m } \right)} \left( {\det \left( \boldsymbol{U} \right)} \right)^{N_t-1 } \Gamma ^{N_t } \left( {M + 1} \right)\Gamma \left( {N_E } \right)\alpha \bar \gamma _B ^{MN_t } }} \nonumber \\
  &\times {}_2F_1\left( {1,MN_t  + N_E ;N_E ;\frac{{N_E \lambda^2_e }}{{\alpha \left( {1 - \lambda^2_e } \right)}}} \right) + o\left( {\frac{1}{{\bar \gamma _B ^{MN_t } }}} \right).
\end{align}
Specially, as $N_E=1$, no combining schemes are deployed at the eavesdropper,
\\${{}_1F_1 \left( {1;N_E ;\frac{{N_E {\lambda^2_e} y}}{{\alpha \bar \gamma _E \left( {1 - \lambda^2_e } \right)^2 }}} \right)}
=\exp\left({\frac{{N_E {\lambda^2_e} y}}{{\alpha \bar \gamma _E \left( {1 - \lambda^2_e } \right)^2 }}}\right)$. The asymptotic SOP of TAS with single antenna eavesdropper is given by
\begin{align}\label{multimrcsinglene}
\mathop {P^{TAS} _{out\_MRC} }\limits_{N_E  = 1} \left( {R_s } \right)=
\frac{{\Gamma \left( {MN_t  + 1} \right)\left( {\bar \gamma _E 2^{R_s } \left( {1 - \lambda ^2_e } \right)} \right)^{MN_t } }}{{\Gamma ^{N_t } \left( {M + 1} \right)\det ^{N_t } \left( U \right)\bar \gamma _B ^{MN_t } }}\left( {\frac{{\sum\limits_{m = 1}^M {\frac{{\eta^2_m }}{{1 - \eta^2_m }} + \frac{{\lambda ^2_e }}{{1 - \lambda ^2_e }}} }}{{\sum\limits_{m = 1}^M {\frac{{\eta^2_m }}{{1 - \eta^2_m }} + 1} }}} \right)^{MN_t }+ o\left( {\frac{1}{{\bar \gamma _B ^{MN_t } }}} \right).
\end{align}
According to \eqref{multimrc2} and \eqref{multimrcsinglene}, the impact of the eavesdropper antenna number $N_E$ on SOP is
\begin{align}\label{impact of number of NE}
\frac{{P^{TAS} _{out\_MRC} \left( {R_s } \right)}}{{\mathop {P^{TAS} _{out\_MRC} }\limits_{N_E  = 1} \left( {R_s } \right)}} = {\textstyle{{\Gamma \left( {MN_t  + N_E } \right)\left( {\sum\limits_{m = 1}^M {\frac{{\eta^2_m }}{{1 - \eta^2_m }} + 1} } \right)^{MN_t  + 1} } \over {\Gamma \left( {N_E } \right)\left( {\sum\limits_{m = 1}^M {\frac{{\eta^2_m }}{{1 - \eta^2_m }} + \frac{{\lambda^2_e}}{{1 - \lambda^2_e}}} } \right)^{MN_t } }}}{}_2F_1\left( {1,MN_t  + N_E ;N_E ;\frac{{N_E \lambda^2_e }}{{\alpha \left( {1 - \lambda^2_e} \right)}}} \right).
\end{align}
It can be easily found that the secrecy array gain decreases with an increase of $N_E$. In addition, since the impact of correlation coefficient on SOP has been analyzed and has the similar result as single-antenna eavesdropper, we do not repeat the analysis.
\subsubsection{SC at the Legitimate Receiver} When SC is applied at the legitimate receiver, we can use the same method as described in Appendix C to obtain the joint PDF of $\gamma^{SC}_{s_{n^*},B}$ and $\gamma^{MRC}_{s_{n^*},E}$. The approximate SOP of TAS/SC is given by
\begin{align}\label{multisc}
&P^{TAS} _{out\_SC} \left( {R_s } \right)= \frac{1}{{\prod\limits_{m = 1}^M {\left( {1 - \eta^2_m } \right)} \det \left( \boldsymbol{U} \right)^{N_t  - 1} \Gamma \left( {N_E } \right)\alpha \bar \gamma _B ^{MN_t } }}\sum\limits_{\omega  = 0}^{MN_t } {\binom{MN_t}{\omega}} \left( {2^{R_s }  - 1} \right)^{MN_t  - \omega }  \nonumber \\
  &\quad \quad\times \Gamma \left( {N_E  + \omega } \right)\left( {2^{R_s } \bar \gamma _E \left( {1 - \lambda^2_e} \right)} \right)^\omega  {}_2F_1\left( {1,N_E  + \omega ;N_E ;\frac{{N_E \lambda^2_e}}{{\alpha \left( {1 - \lambda^2_e} \right)}}} \right) + o\left( {\frac{1}{{\bar \gamma _B ^{MN_t } }}} \right).
\end{align}
As $\eta_m=0$ and $\lambda_e=0$, eq. \eqref{multisc} is reduced to \cite[eq. (33)]{NYang2} over Rayleigh fading. This demonstrates the generality for our result.
\subsubsection{EGC at the Legitimate Receiver} When EGC is applied at the legitimate receiver, we can also use the same method as described in Appendix C to obtain the joint PDF of $\gamma^{EGC}_{s_{n^*},B}$ and $\gamma_{s_{n^*},E}$. The approximate SOP of TAS/EGC is given by
\begin{align}\label{multiegc}
P^{TAS} _{out\_EGC} \left( {R_s } \right)&=\frac{{\left( {2M} \right)^{MN_t }}}{{ {\Gamma ^{N_t }\left( {2M+1} \right)} \prod\limits_{m = 1}^M {\left( {1 - \eta^2_m } \right)} \det^{N_t  - 1} \left( \boldsymbol{U} \right)\Gamma \left( {N_E } \right)\alpha \bar \gamma _B ^{MN_t } }}\nonumber \\
 &\times \sum\limits_{\omega  = 0}^{MN_t } {\binom{MN_t}{\omega}} \Gamma \left( {N_E  + \omega } \right)\left( {2^{R_s }  - 1} \right)^{MN_t  - \omega }\left( {2^{R_s } \bar \gamma _E \left( {1 - \lambda^2_e} \right)} \right)^\omega \nonumber \\
  &\times {}_2F_1\left( {1,N_E  + \omega ;N_E ;\frac{{N_E \lambda^2_e}}{{\alpha \left( {1 - \lambda^2_e} \right)}}} \right) + o\left( {\frac{1}{{\bar \gamma _B ^{MN_t } }}} \right).
\end{align}
\emph{Remark 4:} From \eqref{multimrc}, \eqref{multisc}, and \eqref{multiegc}, the diversity order for three combining schems is $N_tM$. Moreover, $P^{TAS} _{out\_MRC} \left( {R_s } \right)>P^{TAS} _{out\_EGC} \left( {R_s } \right)>P^{TAS} _{out\_SC} \left( {R_s } \right)$. Also $P^{TAS} _{out\_MRC} \left( {R_s } \right)/P^{TAS} _{out\_EGC} \left( {R_s } \right)=\frac{{\left( {\Gamma \left( {2M+1} \right)} \right)^{N_t } }}{{\left( {2M} \right)^{MN_t } \Gamma ^{N_t } \left( {M + 1} \right)}}$ and  $P^{TAS} _{out\_MRC} \left( {R_s } \right)/P^{TAS} _{out\_SC} \left( {R_s } \right)=\frac{1}{{\Gamma ^{N_t } \left( {M + 1} \right)}}$. The gap for different combining schemes does not depend on the eavesdropper's antenna number $N_E$. Especailly, as $N_t=1$ and $N_E=1$,  eqs. \eqref{multimrc}, \eqref{multisc}, and \eqref{multiegc} are simplified to (\ref{asyallcMRC}), (\ref{asyallcsc}), and (\ref{asyegc}). As $\lambda_e=1$, eqs. \eqref{multimrc}, \eqref{multisc}, and \eqref{multiegc} are simplified to (\ref{mrc2}), (\ref{sc2}), and (\ref{EGCc2}).
\subsection{With Eavesdropper's CSI}
When the eavesdropper's CSI is available to the transmitter, the transmit antenna having the largest secrecy capacity is selected, and it can be expressed as \cite{FSAlQahtani2,CKundu}
\begin{align}\label{protcol2}\
n^*  = \arg \mathop {\max }\limits_{1 \le n \le N_t } {\log _2 \left( {\frac{{1 + \gamma _{s_n ,B} }}{{1 + \gamma _{s_n ,E} }}} \right)}.
\end{align}
Though secrecy outage may be avoided when the eavesdropper's CSI is available and the transmitter can adaptively
adjust the transmission parameter such as transmission rate, such implementation
can significantly increase the system complexity. Therefore, the transmitter
may still adopt the constant rate transmission scheme, and secrecy outage can still occur.
\subsubsection{MRC at the Legitimate Receiver}
Since the antenna is selected according to the secrecy capacity, and the SNRs at the legitimate receiver from different antenna are independent, the asymptotic SOP of TAS/MRC is given by
\begin{align}\label{MIMOMRC2}\
P^{TAS'} _{out\_MRC} \left( {R_s } \right)= \prod\limits_{n = 1}^{N_t } {\Pr \left( {C_{s_n }  \le R_s } \right)}
\end{align}
where ${\Pr \left( {C_{s_n }  \le R_s } \right)}$ denotes the SOP of random TAS, and it can be calculated as
\begin{align}\label{MIMOMRCmultime}\
 \Pr &\left( {C_{s_n }  \le R_s } \right)= \int_0^\infty  {dy\int_0^{2^{R_s } \left( {y + 1} \right) - 1} {f_{\gamma _{s_n ,B} ,\gamma _{s_n ,E} } \left( {x,y} \right)} } dx \nonumber \\
  &= \frac{1}{{\prod\limits_{m = 1}^M {\left( {1 - \eta^2_m } \right)} \Gamma \left( {M + 1} \right)\Gamma \left( {N_E } \right)\bar \gamma _B ^M \alpha }}\sum\limits_{\omega  = 0}^M {\binom{M}{\omega}} \left( {2^{R_s }  - 1} \right)^{M - \omega }  \nonumber \\
  &\times \Gamma \left( {N_E  + \omega } \right)\left( {\bar \gamma _E 2^{R_s } \left( {1 - \lambda^2_e} \right)} \right)^\omega  {}_2F_1\left( {1,N_E  + \omega ;N_E ;\frac{{N_E \lambda^2_e}}{{\alpha \left( {1 - \lambda^2_e} \right)}}} \right) + o\left( {\bar \gamma _B ^{ - M} } \right).
\end{align}
Substituting \eqref{MIMOMRCmultime} into \eqref{MIMOMRC2}, the asymptotic SOP of TAS/MRC with eavesdropper's CSI is
\begin{align}\label{MIMOMRC2}\
P^{TAS'} _{out\_MRC} \left( {R_s } \right)&=\frac{1}{{\left( {\prod\limits_{m = 1}^M {\left( {1 - \eta^2_m } \right)} } \right)^{N_t } \Gamma ^{N_t } \left( {M + 1} \right)\Gamma ^{N_t } \left( {N_E } \right)\bar \gamma _B ^{MN_t } \alpha ^{N_t } }} \nonumber\\
  &\times \left( {\sum\limits_{\omega  = 0}^M {\binom{M}{\omega}} \left( {2^{R_s }  - 1} \right)^{M - \omega } \Gamma \left( {N_E  + \omega } \right)\left( {\bar \gamma _E 2^{R_s } \left( {1 - \lambda^2_e} \right)} \right)^\omega  } \right. \nonumber \\
 &\left. { \times {}_2F_1\left( {1,N_E  + \omega ;N_E ;\frac{{N_E \lambda^2_e}}{{\alpha \left( {1 - \lambda^2_e} \right)}}} \right)} \right)^{N_t }  + o\left( {\bar \gamma _B ^{ - MN_t } } \right).
\end{align}
As $N_E=1$, eq. \eqref{MIMOMRC2} can be simplified as
\begin{align}\label{MIMOMRC21}\
&P^{TAS'} _{out\_MRC} \left( {R_s } \right)= \frac{1}{{\left( {\det \left( \boldsymbol{U} \right)} \right)^{N_t } }}\left( {\frac{{\bar \gamma _E }}{{\bar \gamma _B }}} \right)^{MN_t } \nonumber \\
 &\quad \quad \times \left[ {\sum\limits_{k = 0}^M {\frac{1}{{\left( {M - k} \right)!}}\left( {\frac{{2^{R_s } \left( {\sum\limits_{m = 1}^M {\frac{{\eta^2_m \left( {1 - \lambda^2_e } \right)}}{{1 - \eta^2_m }} + 1} } \right)}}{{\left( {\sum\limits_{m = 1}^M {\frac{{\eta^2_m }}{{1 - \eta^2_m }} + 1} } \right)}}} \right)^k\left( {\frac{{2^{R_s }  - 1}}{{\bar \gamma _E }}} \right)^{M - k} } } \right]^{N_t }+ o\left( {\bar \gamma _B ^{ - MN_t } } \right).
\end{align}
From \eqref{MIMOMRC2} and \eqref{MIMOMRC21}, we can easily find the secrecy diversity order is $MN_t$, which does not depend on the eavesdropper antenna number $N_E$.  To further investigate the impact of $N_E$ on the SOP, with $\bar\gamma_E\gg1$,  eq. \eqref{MIMOMRC2} can be rewritten as
\begin{align}\label{MIMOMRC3}\
  P^{TAS'}_{out\_MRC} \left( {R_s } \right)&= \frac{{\Gamma ^{^{N_t } } \left( {N_E  + M} \right)2^{R_s MN_t } \left( {\bar \gamma _E \left( {1 - \lambda^2_e} \right)} \right)^{MN_t } }}{{\left( {\prod\limits_{m = 1}^M {\left( {1 - \eta^2_m } \right)} } \right)^{N_t } \Gamma ^{^{N_t } } \left( {M + 1} \right)\Gamma ^{{N_t } } \left( {N_E } \right)\bar \gamma _B ^{MN_t } \alpha ^{N_t } }} \nonumber \\
  &\times \left( {{}_2F_1\left( {1,N_E  + M;N_E ;\frac{{N_E \lambda^2_e}}{{\alpha \left( {1 - \lambda^2_e} \right)}}} \right)} \right)^{N_t }  + o\left( {\bar \gamma _B ^{ - MN_t } } \right)
\end{align}
which will be used to discuss the gap of different TAS schemes.
\subsubsection{SC at the Legitimate Receiver} Similar to MRC, according to (\ref{asyallcsc}), the asymptotic SOP of TAS/SC is given by
\begin{align}\label{MIMOSC2}\
&P^{TAS'} _{out\_SC} \left( {R_s } \right)
  = \left( {\frac{1}{{\left( {\prod\limits_{m = 1}^M {\left( {1 - \eta^2_m } \right)} } \right)\Gamma \left( {N_E } \right)\bar \gamma _B ^M \alpha }}} \right)^{N_t } \left( {\sum\limits_{\omega  = 0}^M {\binom{M}{\omega }} \left( {2^{R_s }  - 1} \right)^{M - \omega } } \Gamma \left( {N_E  + \omega } \right)\right. \nonumber \\
 &\quad \quad \left. { \times\left( {\bar \gamma _E 2^{R_s } \left( {1 - \lambda^2_e} \right)} \right)^\omega  {}_2F_1\left( {1,N_E  + \omega ;N_E ;\frac{{N_E \lambda^2_e}}{{\alpha \left( {1 - \lambda^2_e} \right)}}} \right)} \right)^{N_t }  + o\left( {\bar \gamma _B ^{ - MN_t } } \right).
\end{align}
\subsubsection{EGC at the Legitimate Receiver} Similar to MRC, according to (\ref{asyegc}), the asymptotic SOP of TAS/EGC is given by
\begin{align}\label{MIMOEGC2}\
P^{TAS'} _{out\_EGC} \left( {R_s } \right)&=
 \left( {\frac{{\left( {2M} \right)^M }}{{\left( {\prod\limits_{m = 1}^M {\left( {1 - \eta^2_m } \right)} } \right)\Gamma \left( {2M+1} \right)\Gamma \left( {N_E } \right)\bar \gamma _B ^M \alpha }}} \right)^{N_t }  \nonumber \\
  &\times \left( {\sum\limits_{\omega  = 0}^M {\binom{M}{\omega }} \left( {2^{R_s }  - 1} \right)^{M - \omega } \Gamma \left( {N_E  + \omega } \right)\left( {\bar \gamma _E 2^{R_s } \left( {1 - \lambda^2_e} \right)} \right)^\omega  } \right. \nonumber \\
 &\left. { \times {}_2F_1\left( {1,N_E  + \omega ;N_E ;\frac{{N_E \lambda^2_e}}{{\alpha \left( {1 - \lambda^2_e} \right)}}} \right)} \right)^{N_t }  + o\left( {\bar \gamma _B ^{ - MN_t } } \right).
\end{align}
As $N_t=1$, eqs. \eqref{MIMOMRC2}, \eqref{MIMOSC2}, and \eqref{MIMOEGC2} specialize to \eqref{multimrc2}, \eqref{multisc}, and \eqref{multiegc}, respectively. Note that when $\eta_m=1$, and $\lambda_e=1$, all the channels are fully correlated. In this case, the SOP of TAS/MRC is given by
\begin{align}\label{MIMOEGC2}\
{P^{TAS'}_{out}} = {\left( {1 - \exp \left( { - \frac{{{2^{{R_s}}} - 1}}{{{{\varepsilon\bar \gamma }_B} - {2^{{R_s}}}N_E{{\bar \gamma }_E}}}} \right)} \right)^{{N_t}}}
\end{align}
where $\varepsilon$ is equal to $M$ for EGC and MRC, and equal to 1 for SC. The inequality $\varepsilon{{\bar \gamma }_B} > {2^{{R_s}}}N_E{{\bar \gamma }_E}$ is required for \eqref{MIMOEGC2}; otherwise, secrecy can not be guaranteed.
\subsection{Discussion}
TAS with eavesdropper's CSI requires the transmitter to obtain instantaneous CSI for all channels, but TAS without eavesdropper's CSI only requires the main channels instantaneous CSI. In addition, it can be seen from (\ref{multimrc})-(\ref{MIMOEGC2}) that with or without eavesdropper's CSI, regardless which combining scheme is deployed, the secrecy diversity order is $M{N_t}$. The difference lies in the secrecy array gain. As $\bar\gamma_{E}\gg 1$,  the ratio of (\ref{multimrc2}) and (\ref{MIMOMRC3}) can be written as
\begin{align}\label{ratio1}\
\frac{{P^{TAS} _{out\_MRC} \left( {R_s } \right)}}{{P^{TAS'} _{out\_MRC} \left( {R_s } \right)}} &=
\frac{{\Gamma \left( {MN_t  + N_E } \right)\Gamma^{N_t-1} \left( {N_E } \right)}}{{\Gamma ^{^{N_t } } \left( {M +N_E } \right)}}\left( {\frac{{\left( {1 + \sum\limits_{m = 1}^M {\frac{{\eta^2_m }}{{1 - \eta^2_m }}}  + \frac{{N_E \lambda^2_e}}{{1 - \lambda^2_e}}} \right)}}{{\left( {1 + \sum\limits_{m = 1}^M {\frac{{\eta^2_m }}{{1 - \eta^2_m }}} } \right)}}} \right)^{N_t  - 1}\nonumber \\
&\times \frac{{{}_2F_1\left( {1,MN_t  + N_E ;N_E ;\frac{{N_E \lambda^2_e}}{{\alpha \left( {1 - \lambda^2_e} \right)}}} \right)}}{{\left( {{}_2F_1\left( {1,M + N_E ;N_E ;\frac{{N_E \lambda^2_e}}{{\alpha \left( {1 - \lambda^2_e} \right)}}} \right)} \right)^{N_t } }}.
\end{align}
In \eqref{ratio1}, the first part can be expressed as
\begin{align}\label{ratio1part}\
\frac{{\Gamma \left( {MN_t  + N_E } \right)\Gamma ^{N_t  - 1} \left( {N_E } \right)}}{{\Gamma ^{^{N_t } } \left( {M + N_E } \right)}} = \frac{{\underbrace {\left( {MN_t  + N_E  - 1} \right) \cdots \left( {N_E  + 1} \right)N_E }_{MN_t } \times \Gamma ^{N_t } \left( {N_E } \right)}}{{\left( {\underbrace {\left( {M + N_E  - 1} \right) \cdots \left( {N_E  + 1} \right)N_E }_M} \right)^{^{N_t } }  \times \Gamma ^{^{N_t } } \left( {N_E } \right)}} \geq 1.
\end{align}
Note in \eqref{ratio1part} as $N_t=1$, the equality is satisfied.
 Moreover, the second part of \eqref{ratio1},\\
 $\left( {\frac{{1 + \sum\limits_{m = 1}^M {\frac{{\eta^2_m }}{{1 - \eta^2_m }}}  + \frac{{N_E \lambda^2_e}}{{1 - \lambda^2_e}}}}{{1 + \sum\limits_{m = 1}^M {\frac{{\eta^2_m }}{{1 - \eta^2_m }}} }}} \right)^{N_t  - 1}$, is also larger than $1$, and the third part $\frac{{{}_2F_1\left( {1,MN_t  + N_E ;N_E ;\frac{{N_E \lambda^2_e}}{{\alpha \left( {1 - \lambda^2_e} \right)}}} \right)}}{{\left( {{}_2F_1\left( {1,M + N_E ;N_E ;\frac{{N_E \lambda^2_e}}{{\alpha \left( {1 - \lambda^2_e} \right)}}} \right)} \right)^{N_t } }} \simeq 1$. Obviously, the gap is increased with $N_t$. For the special case of $N_E=1$, eq. \eqref{ratio1} is simplified to
 \begin{align}\label{ratio2}\
\frac{{P^{TAS} _{out\_MRC} \left( {R_s } \right)}}{{P^{TAS'} _{out\_MRC} \left( {R_s } \right)}} = \frac{{\left( {MN_t } \right)!}}{{\left( {M!} \right)^{N_t} }}
\end{align}
where the identity ${}_2F_1\left( { - n,1;1; - z} \right) = \left( {1 + z} \right)^n$ is used \cite[eq. (9.121.1)]{IS} to obtain \eqref{ratio2}. Similar to \eqref{ratio1part}, we have $\left( {MN_t } \right)! \ge \left( {M!} \right)^{N_t }$. Then $P^{TAS} _{out\_MRC} \left( {R_s } \right) \ge P^{TAS'} _{out\_MRC} \left( {R_s } \right)$. In addition, with SC and EGC at the legitimate receiver, using \eqref{multisc}, \eqref{multiegc}, \eqref{MIMOSC2}, and \eqref{MIMOEGC2}, similar analytical results can be obtained. Due to space limitation, we omit the analyse here.

\emph{Remark 5}: The SOP of TAS with eavesdropper's CSI outperforms the case without eavesdropper's CSI, regardless which combining scheme is deployed at the legitimate receiver. As $\bar\gamma_{B}\rightarrow \infty$, the SOP gap with eavesdropper CSI and without eavesdropper CSI is increased with $N_t$, and it has no relationship with $N_E$.
\section{Simulation Results}
In this section, we present numerical and simulation results
to verify our analysis. All the links are assumed to
experience Rayleigh flat fading. Without loss of generality, we set $R_s=1$ bit/sec/Hz.

When the multi-antenna main channels are correlated and independent of the eavesdropper channel, the SOPs of MRC, SC, and EGC are plotted in Fig. 2, where $\boldsymbol{U}_1$  denotes the correlation matrix at the multi-antenna legitimate receiver, and it is given by
\begin{equation}\boldsymbol{U}_1=\left(\begin{array}{ccc}
1 & 0.765  &  -0.8075\\
0.765  & 1  &  -0.8550 \\
 -0.8075 &  -0.8550 & 1
 \end{array}\right)\end{equation}
where $\det({\boldsymbol{U}_1})=0.088$. In $\boldsymbol{U}_1$, the coefficients $\eta_1=0.8, \eta_2=0.85, \eta_3=-0.95$. $\boldsymbol{U}_2$ denotes the case that the main channels are completely correlated, namely $\eta_m=1$. The curves named independent in Fig. 2 represent the case when all the channels are independent. We can find that compared to the case where main channels are independent, CMC degrades the SOP. This is because when the main channels are independent, i.e. $\eta_{m}=0$, the determinant of correlation matrix is 1, and is greater than $\det({\boldsymbol{U}_1})$. Namely, antenna correlation at the legitimate receiver has a negative effect on SOP, as shown in Remark 1. In addition, we can observe that MRC is better than EGC and SC except for the completely correlated case. It is because when the main channels are completely correlated, only one effective channel plays a role for diversity reception. Thus, for the completely correlated case, SC, EGC and MRC have the same SOP performance. Also, it can be seen that simulation results coincide with the analytical results. When $\bar\gamma_B$ is increased, the asymptotic results converge to the simulation and analytical results, which verifies our analysis.

\begin{figure}
\centering
\includegraphics[width=0.48\textwidth]{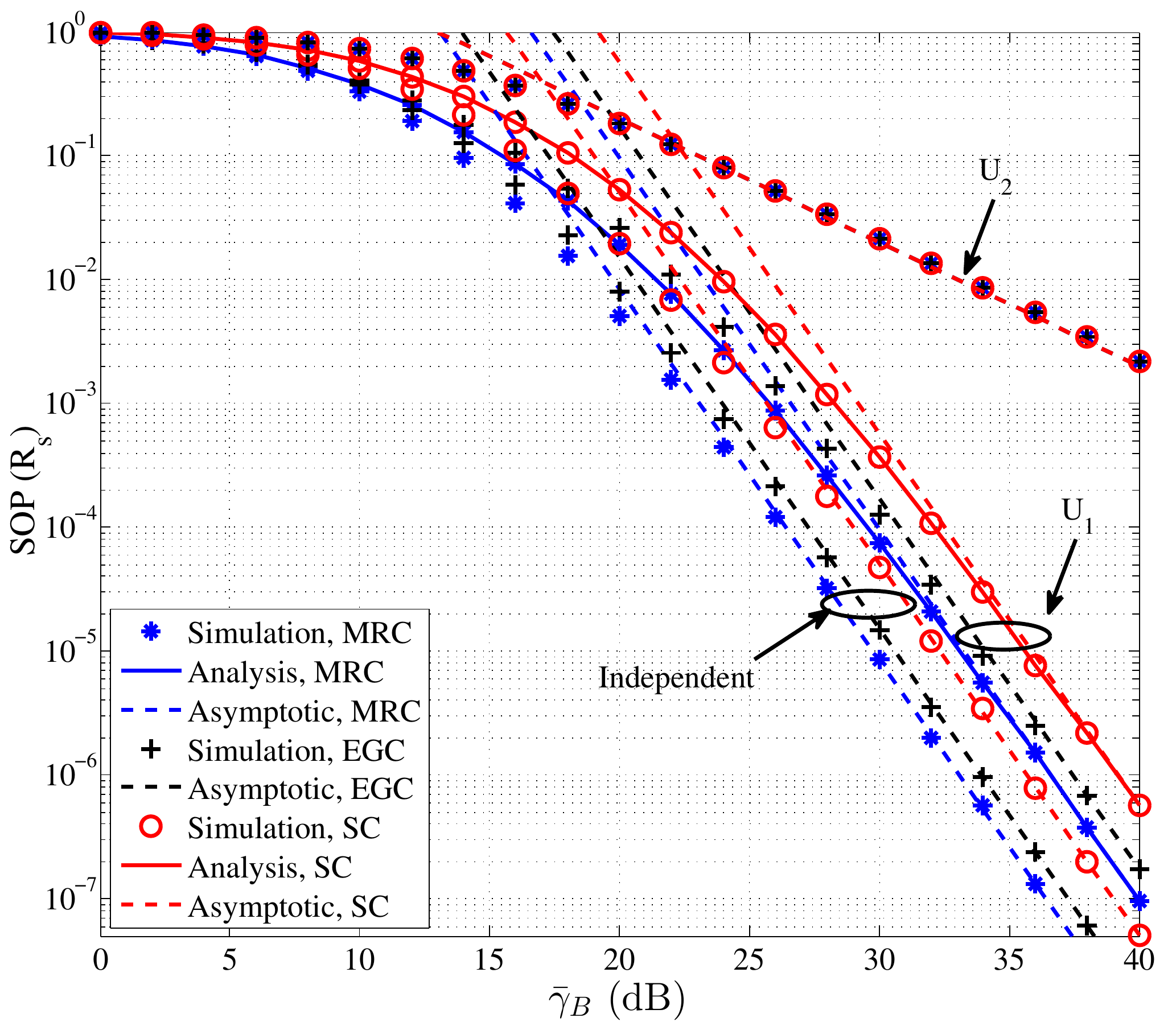}
\caption{The SOP  versus $\bar\gamma_{B}$ under the case of CMC. $M=3$, $N_t=1$, $N_E=1$, $\lambda_e=0$, and $\bar\gamma_{E}=10$ dB.} \label{Fig2}
\end{figure}
\begin{figure}
\centering
\includegraphics[width=0.48\textwidth]{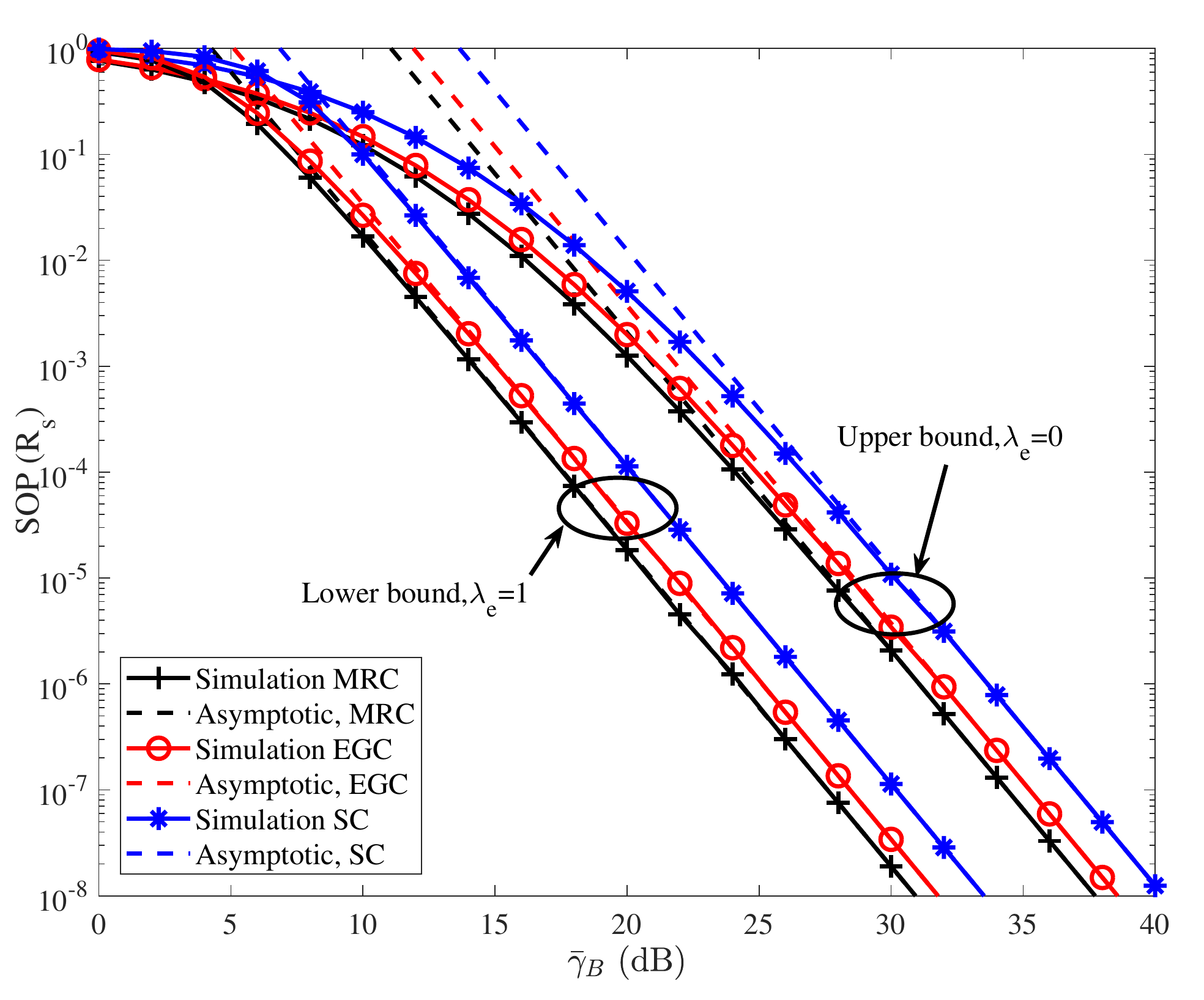}
\caption{The lower and upper bounds of SOP versus $\bar\gamma_{B}$. $M=3$, $N_t=1$, $N_E=1$, and $\bar\gamma_{E}=5$ dB.} \label{Fig3}
\end{figure}
\begin{figure}
\centering
\includegraphics[width=0.48\textwidth]{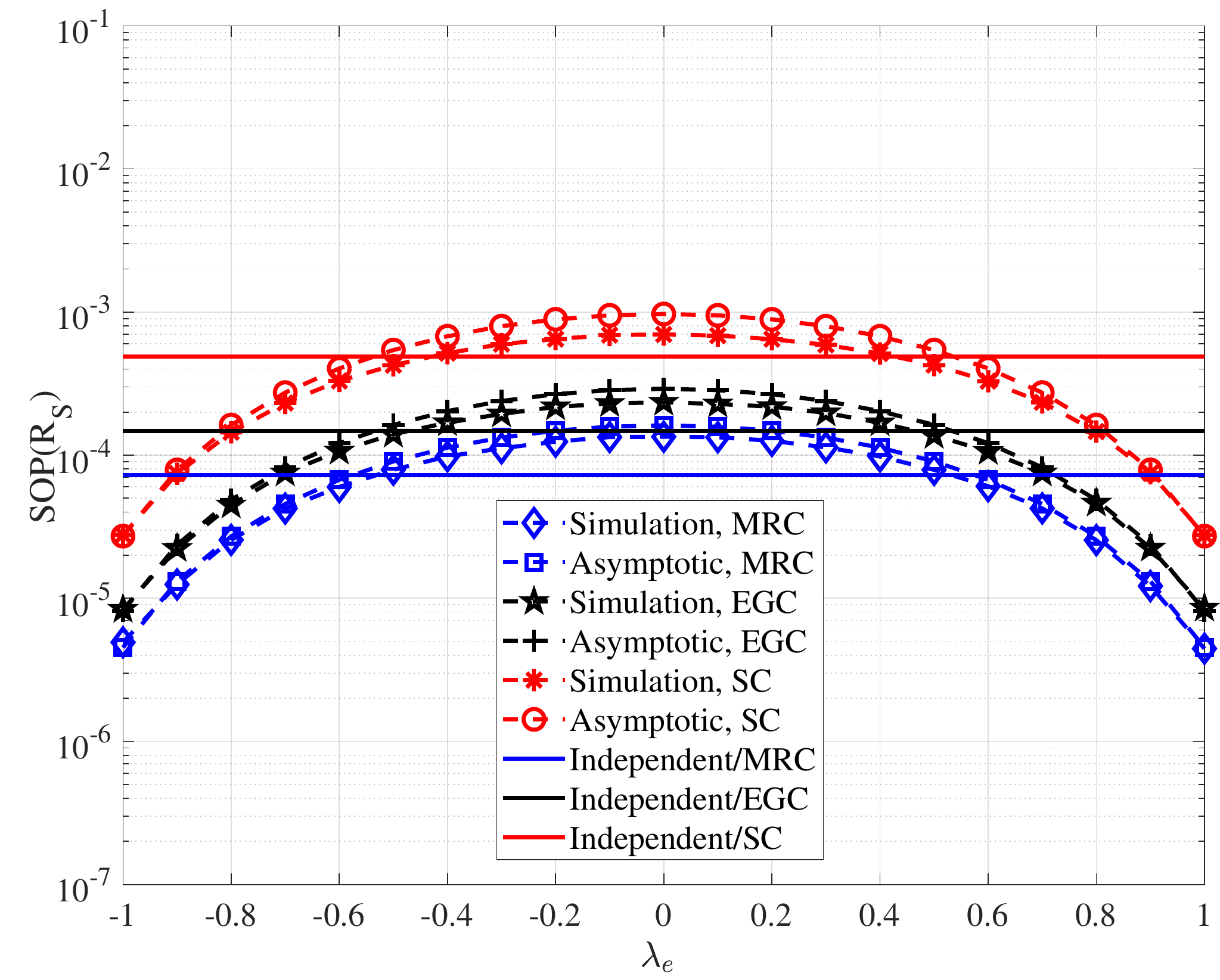}
\caption{The impact of $\lambda_e$ on SOP. $M=3$, $N_t=1$, $\bar\gamma_{B}=20$ dB, and $\bar\gamma_{E}=3$ dB.} \label{Fig4}
\end{figure}
The lower and upper bounds of SOP for MRC, SC, and EGC are illustrated in Fig. 3. The main channel correlation matrix is $\boldsymbol{U}_1$.  We can observe that the SOP with $\lambda_e=1$ is clealy better than that with $\lambda_e=0$, which verifies our prediction that the correlation between the main channels and eavesdropper channels has a positive impact on SOP when the average SNR at the legitimater receiver is larger than that at the eavesdropper. When $\bar\gamma_{B}$ is increased, the simulation results converge to the asymptotic results. In addition, it can be found that all the asymptotic curves have the same slope. This is because all the combining schemes have the same secrecy diversity order $M$, as predicted by our analysis.

The impact of $\lambda_e$ on SOP is examined in Fig. 4, where the dashed line denotes the scenario when all the channels are correlated and the main channels correlation matrix is $\boldsymbol{U}_3$; the solid line denotes the scenario all channels are independent. $\boldsymbol{U}_3$ is given by
 \begin{equation}\boldsymbol{U}_3=\left(\begin{array}{ccc}
1 & -0.42  &  0.48 \\
-0.42  & 1  &  -0.56 \\
 0.48 &  -0.56 & 1
 \end{array}\right).\end{equation}
In $\boldsymbol{U}_3$, the correlation coefficients are $\eta_1=0.6, \eta_2=-0.7, \eta_3=0.8$, and $\det({\boldsymbol{U}_3})= 0.5054$. It can be seen that as $\lambda_e$ is small, the SOP performance of the scenario when all the channels are independent is better than the scenario when all the channels are correlated. But as $\lambda_e$ is increased, the contrary result is obtained. In other words, the SOP when all the channels are correlated can be lower than that when all the channels are independent, which agree with the discussion in Remark 3.
\begin{figure}
\centering
\includegraphics[width=0.45\textwidth]{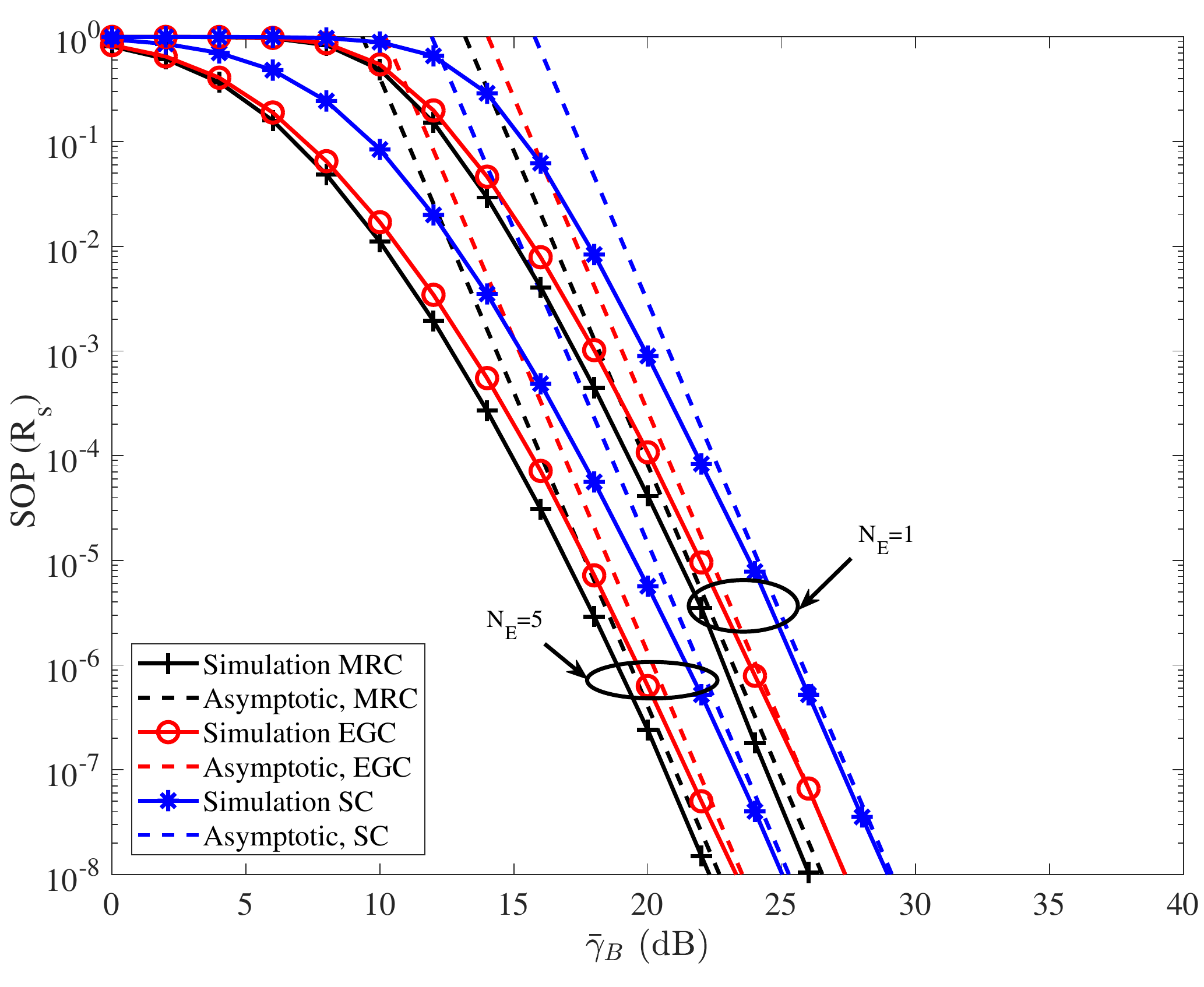}
\caption{The SOP of TAS without the eavesdropper's CSI. $\lambda_{e}=0.8$, $N_{t}=2$, and $\bar\gamma_{E}=5$ dB.} \label{Fig5}
\end{figure}
\begin{figure}
\centering
\includegraphics[width=0.45\textwidth]{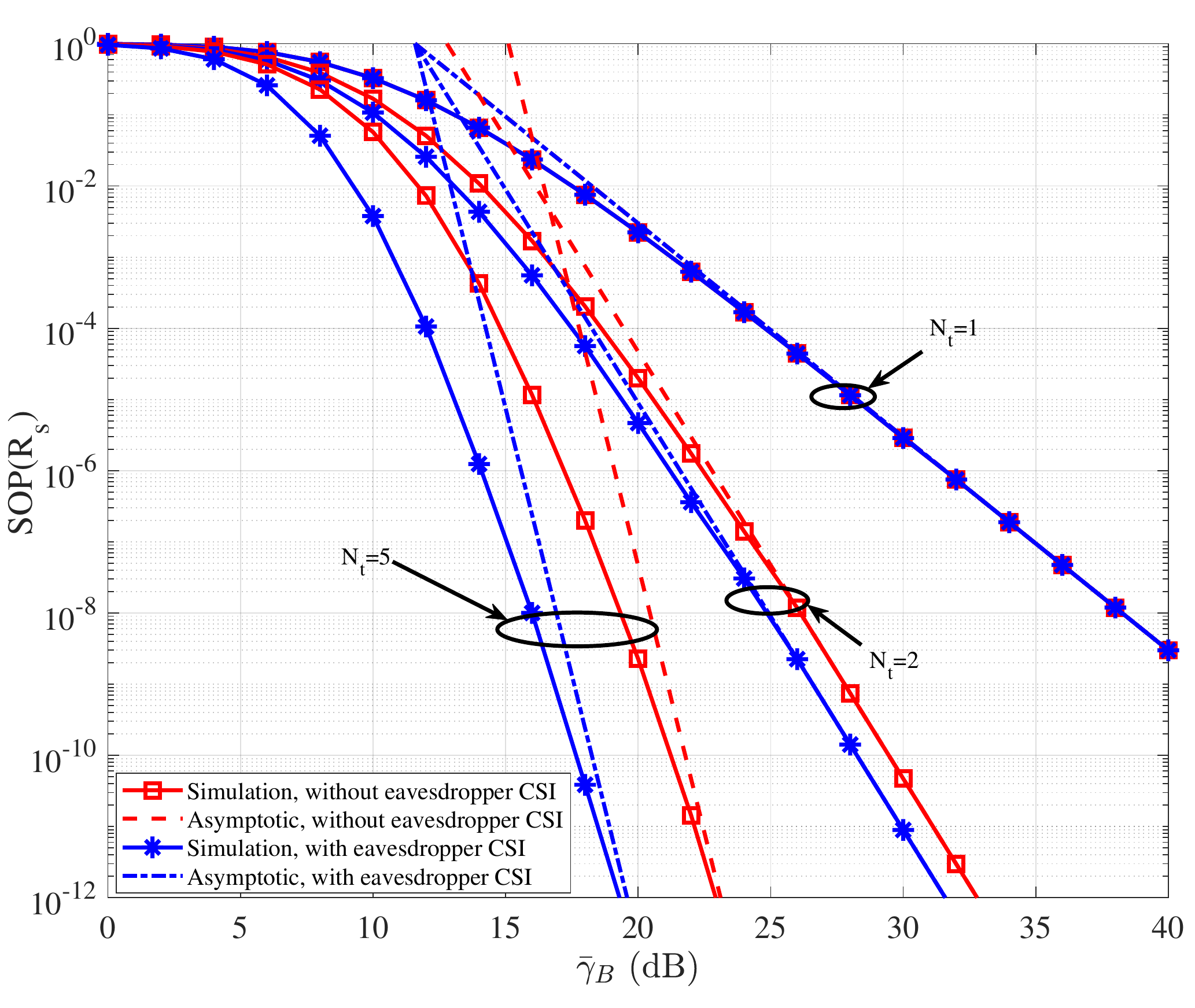}
\caption{The SOP of TAS/MRC with or without the eavesdropper's CSI. $M=3$, $N_E=3$, $\lambda_{e}=0.5$, and $\bar\gamma_{E}=5$ dB.} \label{Fig6}
\end{figure}

For different number of antennas at the eavesdropper,  with or without the eavesdropper's CSI, the SOPs of TAS with MRC, SC, and EGC are illustrated in Fig. 5, where the main channel correlation matrix is $\boldsymbol{U}_1$. The SOP is increased when the number of antennas at the eavesdropper is increased, which verifies our secrecy analysis in Section IV. Also, we can observe all the curves have the same slope. It is because no matter with or without eavesdropper's CSI, the secrecy diversity order is ${N_t}M$ and has no relationship with $N_E$.

When multiple antennas are deployed at the transmitter and the eavesdropper, the SOPs of TAS/MRC with or without eavesdropper's CSI are plotted in Fig. 6, where the main channels have correlation matrix $\boldsymbol{U}_3$. It can be seen that the SOP of TAS with eavesdropper's CSI outperforms that without eavesdropper's CSI. This is because TAS with eavesdropper selects the antenna according to the maximum secrecy capacity.  Also, we can find that as $N_t=1$, TAS with eavesdropper's CSI has the same SOP as TAS without eavesdropper's CSI as expected. The gap of SOP for the two TAS cases is increased with $N_t$, which verifies our analysis in Remark 5.
\section{Conclusion}
We considered a comprehensive secure communication system model with generalized correlation structure. The impacts of both antenna correlation and channel correaltion between the legitimate receiver and eavesdropper on the SOP were investigated and quantified in detail. The findings suggest that the antenna correlation at the legitimate receiver alone is detrimental to the SOP. However, when the average SNR at the legitimate is greater than that at the eavesdropper, the correlation between the main and eavesdropper's channels can enhance the SOP. Finally, for an eavesdropper with multiple antennas, the gaps of SOP for TAS with and without eavesdropper's CSI increase with the number of transmitter antennas.
\begin{appendices}
\section{Approximate Conditional PDF of $\gamma^{MRC} _{s_n,B}$}
Assuming ${T=X^2_{n,0}+Y^2_{n, 0}}$, $X_{n,0}=x_0$, and $Y_{n,0}=y_0$. Conditioned on $x^2_0+y^2_0=t$, the MGF of $\gamma^{MRC} _{s_n,B}$ can be expressed as \cite{AAnnamalai}
\begin{align}\label{conmgfMRC}\
{\phi _{\left. {{\gamma^{mrc} _{s_n,B}}} \right|T}}\left( {s\left| t \right.} \right) &= E\left( {\left. {\exp \left( { - s\sum\limits_{m = 1}^M {{\gamma _{s_n,{b_m}}}} } \right)} \right|t} \right) = \prod\limits_{m = 1}^M {E\left( {\left. {\exp \left( { - s{\gamma _{s_n,{b_m}}}} \right)} \right|t} \right)} \nonumber \\
 &= \prod\limits_{m = 1}^M {{\phi _{\left. {{\gamma_{s_n,b_m}}} \right|T}}\left( {s\left| t \right.} \right)}
\end{align}
where ${{\phi _{\left. {{\gamma_{s_n,b_m}}} \right|T}}\left( {s\left| t \right.} \right)}$ denotes MGF of ${\gamma_{s_n,b_m}}$ conditioned on $T$. Note that the conditional PDF of $\gamma_{s_n,b_m}$ can be obtained by differentiating (\ref{CDFsbk}) with respect to $y$, and it is given by
\begin{align}\label{PDFsbk}\
{f_{{\gamma _{s_n,b_m}}|T}}\left( {y\left| t \right.} \right) = \frac{{\exp \left( { - \frac{{{\eta^2_m}t}}{{1 - {\eta^2_m}}}} \right)}}{{{\bar\gamma _B}\left( {1 - {\eta^2_m}} \right)}}\exp \left( { - \frac{y}{{{\bar\gamma _B}\left( {1 - {\eta^2_m}} \right)}}} \right){I_0}\left( {2\sqrt {\frac{{{\eta^2_m}t y}}{\bar\gamma_B({1 - {\eta^2_m}})^2}}} \right).
\end{align}
According to (\ref{PDFsbk}), the MGF of ${\gamma_{s_n,b_m}}$ conditioned on $T$ can be calculated as
\begin{align}\label{conmgfsigle}\
&{\phi _{\left. {{\gamma_{s_n,b_m}}} \right|T}}\left( {s\left| t \right.} \right) = \int_0^\infty  {{e^{ - sy}}} {f_{{\gamma _{s_n,{b_m}|T}}}}\left( {\left. y \right|t} \right)dy \nonumber \\
 &= \frac{{\exp \left( { - \frac{{{\eta^2_m}t}}{{1 - {\eta^2_m}}}} \right)}}{{{{\bar \gamma }_B}\left( {1 - {\eta^2_m}} \right)}}\int_0^\infty  {\exp \left( { - \left( {s + \frac{1}{{{{\bar \gamma }_B}\left( {1 - {\eta^2_m}} \right)}}} \right)y} \right)} {I_0}\left( {2\sqrt {\frac{{{\eta^2_m}ty}}{{{{\bar \gamma }_B}{{\left( {1 - {\eta^2_m}} \right)}^2}}}} } \right)dy \nonumber \\
 &= \frac{{\exp \left( { - \left( {\frac{{{\eta^2_m}{{\bar \gamma }_B}ts}}{{\left( {1 + s{{\bar \gamma }_B}\left( {1 - {\eta^2_m}} \right)} \right)}}} \right)} \right)}}{{{{\bar \gamma }_B}\left( {1 - {\eta^2_m}} \right)s + 1}}.
\end{align}
As $\bar\gamma_{B}\rightarrow \infty$, $s{{\bar \gamma }_B}\left( {1 - {\eta^2_m}} \right) \gg 1$. Thus the asymptotic conditional MGF of ${\gamma_{s_n,b_m}}$ is
\begin{align}\label{asyconmgfsigle}\
{\phi _{\left. {{\gamma_{s_n,b_m}}} \right|T}}\left( {s\left| t \right.} \right) \simeq \frac{{\exp \left( { - \left( {\frac{{{\eta^2_m}t}}{{\left( {1 - {\eta^2_m}} \right)}}} \right)} \right)}}{{{{\bar \gamma }_B}\left( {1 - {\eta^2_m}} \right)s}}.
\end{align}
Then, the asymptotic conditional MGF $\gamma^{MRC}_{s_n,B}$ can be obtained as
\begin{align}\label{asyconmgf}
{\phi _{\left. {{\gamma^{MRC} _{s_n,B}}} \right|T}}\left( {s\left| t \right.} \right) = \frac{{\exp \left( { - {\sum\limits_{m = 1}^M {\frac{{{\eta^2_m}t}}{{\left( {1 - {\eta^2_m}} \right)}}} } } \right)}}{{{{\bar \gamma }_B}^M}}\prod\limits_{m = 1}^M {\frac{1}{{\left( {1 - {\eta^2_m}} \right)s^M}}}.
\end{align}
Performing the  inverse Laplace transform for (\ref{asyconmgf}), we obtain the conditional PDF of $\gamma^{MRC}_{s_n,B}$ as
\begin{align}\label{asyconpdf}
f_{\left. {{\gamma^{MRC}_{s_n,B}}} \right|T}\left( {x\left| t \right.} \right) = {\mathcal{L}^{ - 1}}\left( {{\phi _{\left. {{\gamma^{MRC}_{s_n,B}}} \right|T}}\left( {s\left| t \right.} \right)} \right) = \frac{{\exp \left( { - {\sum\limits_{m = 1}^M {\frac{{{\eta^2_m}t}}{{\left( {1 - {\eta^2_m}} \right)}}} }} \right)\prod\limits_{m = 1}^M {\frac{1}{{\left( {1 - {\eta^2_m}} \right)}}} x^{M - 1}}}{{{{\Gamma \left( M \right)\bar \gamma }_B}^M}}.
\end{align}
\section{Approximate Conditional PDF of $\gamma^{EGC} _{s_n,B}$}
According to (\ref{SNREGC}), to derive the PDF of ${\gamma^{EGC} _{s_n, B} }$, we are required to derive the PDF of $\sqrt {\gamma^{EGC} _{s_n, B} }$. Let $Z = \sqrt {\gamma^{EGC} _{s_n, B} }  = \sum\limits_{m = 1}^M {\sqrt {\frac{{\gamma _{s_n,b_m } }}{M}} }$, conditioned on $x^2_0+y^2_0=t$, the MGF of $Z$  is given by
\begin{align}\label{MGFZ}\
 \phi _{\left. Z \right|T} \left( {s\left| t \right.} \right) &= E\left( {\left. {\exp \left( { - s
\left( \sum\limits_{m = 1}^M {\sqrt {\frac{{\gamma _{s_n,b_m } }}{M}} }\right)
} \right)} \right|t} \right) = \prod\limits_{m = 1}^M {E\left( {\left. {\exp \left( { - s{\sqrt{\frac{{{\gamma_{s_n,b_m }}}}{{M }}}}} \right)} \right|t} \right)}  \nonumber \\
  &= \prod\limits_{m = 1}^M {\phi _{\left. {\sqrt{\frac{{{\gamma_{s_n,b_m }}}}{{M }}}} \right|T} \left( {\left. s \right|t} \right)}
\end{align}
where $
\phi _{\left. {{\sqrt{\frac{{{\gamma_{s_n,b_m }}}}{{M }}}}} \right|T} \left( {\left. s \right|t} \right)
$ denotes the MGF of ${\sqrt{\frac{{{\gamma_{s_n,b_m }}}}{{M }}}}$ conditioned on $T$. According to (\ref{PDFsbk}), performing a variable transformation, and the PDF of ${\sqrt{\gamma_{s_n,b_m }}}$ conditioned on $T$ is given by
\begin{align}\label{PDFrootsbk}\
{f_{{\sqrt{\gamma_{s_n,b_m }}|T}}}\left( {x\left| t \right.} \right) =\frac{{2\exp \left( { - \frac{{{\eta^2_m}t}}{{1 - {\eta^2_m}}}} \right)}}{{{\bar\gamma _B}\left( {1 - {\eta^2_m}} \right)}}x\exp \left( { - \frac{x^2}{{{\bar\gamma _B}\left( {1 - {\eta^2_m}} \right)}}} \right){I_0}\left( {2\sqrt {\frac{{{\eta^2_m}t x^2}}{\bar\gamma_B({1 - {\eta^2_m}})^2}}} \right).
\end{align}
Thus, $\phi _{\left. {{\sqrt{\frac{{{\gamma_{s_n,b_m }}}}{{M }}}}} \right|T} \left( {\left. s \right|t} \right)
$ can be calculated as
\begin{align}\label{MGFZ1}\
 &\phi_{\left. {{\sqrt{\frac{{{\gamma_{s_n,b_m }}}}{{M }}}}} \right|T} \left( {\left. s \right|t} \right) = \int_0^\infty  {\exp \left( { - \frac{{sx}}{{\sqrt M }}} \right)f_{\left. {\sqrt{\gamma_{s_n,b_m }}} \right|T} \left( {\left. x \right|t} \right)} dx \nonumber \\
  &= \int_0^\infty  {\exp \left( { - \frac{{sx}}{{\sqrt M }}} \right)\frac{{2\exp \left( { - \frac{{\eta^2_m t}}{{1 - \eta^2_m }}} \right)}}{{\bar \gamma _B \left( {1 - \eta^2_m } \right)}}x\exp } \left( { - \frac{{x^2 }}{{\bar \gamma _B \left( {1 - \eta^2_m } \right)}}} \right)I_0 \left( {\sqrt {\frac{{\eta^2_m tx^2 }}{{\bar \gamma _B \left( {1 - \eta^2_m } \right)}}} } \right)dx.
\end{align}
As $\bar \gamma_B\rightarrow\infty$, $
I_0 \left( {\sqrt {\frac{{\eta^2_m tx^2 }}{{\bar \gamma _B \left( {1 - \eta^2_m } \right)}}} } \right)\simeq 1
$, and $\exp \left( { - \frac{{x^2 }}{{\bar \gamma _B \left( {1 - \eta^2_m } \right)}}} \right) \simeq 1$. Thus, the MGF of $
{{\sqrt{\frac{{{\gamma_{s_n,b_m }}}}{{M }}}}}
$ conditioned on $T$ is approximated as
\begin{align}\label{asyMGFhbk}\
{\phi _{\left. {{\sqrt{\frac{{{\gamma_{s_n,b_m }}}}{{M }}}}} \right|T} \left( {\left. s \right|t} \right)}  &\simeq \int_0^\infty  {x\exp \left( { - \frac{{sx}}{{\sqrt M }}} \right)\frac{{2\exp \left( { - \frac{{\eta^2_m t}}{{1 - \eta^2_m }}} \right)}}{{\bar \gamma _B \left( {1 - \eta^2_m } \right)}}dx} = \frac{{2 M\exp \left( { - \frac{{\eta^2_m t}}{{1 - \eta^2_m }}} \right)}}{{\bar \gamma _B \left( {1 - \eta^2_m } \right)s^2 }}.
\end{align}
Substituting (\ref{asyMGFhbk}) into (\ref{MGFZ}), we obtain
\begin{align}\label{MGFZ2}\
\phi _{\left. Z \right|T} \left( {\left. s \right|t} \right) \simeq \frac{{\left( {2M} \right)^M \exp \left( { - \sum\limits_{k = 1}^M {\frac{{\eta^2_m t}}{{1 - \eta^2_m }}} } \right)}}{{\bar \gamma _B ^M \prod\limits_{k = 1}^M {\left( {1 - \eta^2_m } \right)s^{2M}} }}.
\end{align}
Performing the inverse Laplace transform of (\ref{MGFZ2}), we obtain the PDF of $Z$ conditioned on $T$ as
\begin{align}\label{MGFZ2}\
f_{\left. Z \right|T} \left( {\left. z \right|t} \right) \simeq \frac{{\left( {2M} \right)^M \exp \left( { - \sum\limits_{k = 1}^M {\frac{{\eta^2_m t}}{{1 - \eta^2_m }}} } \right)z^{2M - 1}}}{{\Gamma \left( {2M} \right)\bar \gamma _B ^M \prod\limits_{k = 1}^M {\left( {1 - \eta^2_m } \right)} }}.
\end{align}
Since $\gamma^{EGC}_{s_n,B}=Z^2$, after a proper transformation, the PDF of $\gamma^{EGC}_{s_n,B}$ conditioned on $T$ is
\begin{align}\label{PDFZ2}\
f_{\left. {\gamma^{EGC} _{s_n, B} } \right|T} \left( {\left. x \right|t} \right) \simeq \frac{{\left( {2M} \right)^M \exp \left( { - \sum\limits_{m = 1}^M {\frac{{\eta^2_m t}}{{1 - \eta^2_m }}} } \right)}x^{M-1}}{{{2\Gamma \left( {2M} \right)}\bar \gamma _B ^M \prod\limits_{m = 1}^M {\left( {1 - \eta^2_m } \right)} }}.
\end{align}
\end{appendices}
\section{Approximate Joint PDF of $\gamma^{MRC} _{s_{n^*,B}}$ and $\gamma^{MRC} _{s_{n^*,E}}$}
The PDF of $\gamma^{MRC} _{s_{n^*,B} }$ is given by
\begin{align}\label{jointsinglepdf}
f_{\gamma^{MRC}_{s_{n^*,B} } } \left( x \right) = N_t \left( {F_{\gamma^{MRC} _{s_{n,B} } } \left( x \right)} \right)^{N_t  - 1} f_{\gamma^{MRC} _{s_{n,B} } } \left( x \right).
\end{align}
Then, the joint PDF of $\gamma^{MRC}_{s_{n^*},B}$ and $\gamma^{MRC}_{s_{n^*},E}$ is calculated as
\begin{align}\label{Jnd}\
f_{ \gamma^{MRC}_{s_{n^*} ,B}, \gamma^{MRC}_{s_{n^*} ,E}} \left( {x,y} \right)&= \frac{{ f_{\gamma^{MRC}_{s_n,B},\gamma^{MRC} _{s_n,E}} \left( {x,y} \right)}}{{f_{\gamma^{MRC}_{s_n,B} } \left( x \right)}}{N_t}\left( {F_{\gamma^{MRC}_{s_n,B} } \left( x \right)} \right)^{{N_t} - 1} f_{\gamma^{MRC}_{s_n,B} } \left( x \right) \nonumber \\
  &={N_t}f_{\gamma^{MRC}_{s_n,B},\gamma ^{MRC}_{s_n,E} } \left( {x,y} \right)\left({F_{\gamma^{MRC}_{s_n,B} } \left( x \right)} \right)^{{N_t} - 1}
\end{align}
where, with the help of \eqref{pdfemrc}, the joint PDF of $\gamma^{MRC}_{s_n,B}$ and $\gamma^{MRC}_{s_n,E}$ can be written as
\begin{align}\label{jointsinglepdf}
&f_{\gamma^{MRC}_{s_n,B},\gamma^{MRC} _{s_n ,E}} \left( {x,y} \right)= \int_0^\infty  {\exp \left( { - t} \right)} f_{\gamma^{MRC}_{s_n ,B}|T} \left( {\left. x \right|t} \right)f_{\gamma^{MRC}_{s_n ,E}|T} \left( {\left. y \right|t} \right)dt \nonumber \\
  &= \int_0^\infty  {\frac{{\prod\limits_{m = 1}^M {\frac{1}{{1 - \eta^2_m }}} }}{{\Gamma \left( M \right)\bar \gamma _B ^M }}x^{M - 1} \frac{1}{{\bar \gamma _E \left( {1 - \lambda^2_e} \right)}}\left( {\frac{y}{{N_E \bar \gamma _E {\lambda^2_e}t }}} \right)^{\frac{{N_E  - 1}}{2}} } \nonumber \\
  &\quad \times \exp \left( { - t - \sum\limits_{m = 1}^M {\frac{{\eta^2_m t}}{{1 - \eta^2_m }} - \frac{{N_E {\lambda^2_e}t}}{{\left( {1 - \lambda^2_e} \right)}}} } \right)I_{N_E  - 1} \left( {2\sqrt {\frac{{N_E {\lambda^2_e}ty}}{{\bar \gamma _E \left( {1 - \lambda^2_e} \right)^2 }}} } \right)dt \nonumber \\
  &= \frac{{ {\left( {\frac{1}{{1 - \lambda^2_e}}} \right)^{N_E }x^{M - 1} y^{N_E  - 1}\exp \left( { \frac{ -y}{{\bar \gamma _E \left( {1 - \lambda^2_e} \right)}}} \right) } }}{{\prod\limits_{m = 1}^M({1 - \eta^2_m })\Gamma \left( M \right)\bar \gamma _B ^M \bar \gamma _E ^{N_E } \Gamma \left( {N_E } \right)\alpha }}{}_1F_1 \left( {1;N_E ;\frac{{N_E {\lambda^2_e}y}}{{\alpha  {\bar \gamma _E \left( {1 - \lambda^2_e} \right)^2 } }}} \right) +
o\left( {\frac{{x^{M-1} }}{{\bar \gamma _B ^{M } }}} \right)
\end{align}
where $\alpha  = 1 + \sum\limits_{m = 1}^M {\frac{{\eta^2_m }}{{1 - \eta^2_m }} + \frac{{N_E \lambda^2_e}}{{\left( {1 - \lambda^2_e} \right)}}}$. Both identities \cite[eq. (6.643.2)]{IS} and \cite[eq. (9.220.2)]{IS} are used for obtaining \eqref{jointsinglepdf}.
Integrating (\ref{asyconpdf}) with respect to (w. r. t) $t$, we can obtain an approximate CDF of $\gamma^{MRC} _{s_{n,B} }$ as
\begin{align}\label{rbmrccdf}
F_{\gamma^{MRC}_{s_{n,B} } } \left( x \right) = \frac{{x^M }}{{M!\det \left( \boldsymbol{U} \right){\bar\gamma_{B}}^M }}+
o\left( {\frac{{x^{M} }}{{\bar \gamma _B ^{M } }}} \right).
\end{align}
Substituting ({\ref{rbmrccdf}}) and ({\ref{jointsinglepdf}}) into (\ref{Jnd}),  the asymptotic joint PDF of $\gamma^{MRC} _{s_{n^*,B}}$ and $\gamma^{MRC} _{s_{n^*,E}}$ can be obtained as \eqref{jointpdf}.

\bibliography{SecrecyOutagePerformance}

\end{document}